\documentclass[twocolumn]{aa}  

\usepackage{longtable}
\usepackage{graphicx}
\usepackage{txfonts}
\usepackage{amssymb}
\usepackage{esvect}
\usepackage{natbib,twoopt}
\usepackage{color}

\usepackage{xcolor}
\begin{document}

%%%%%%%%%%%%%%%%%%%%%%%%%%%%%%%%%%%%%%%%
%\usepackage[options]{hyperref}
% To add links in your PDF file, use the package "hyperref"
% with options according to your LaTeX or PDFLaTeX drivers.
%
   \title{The Fornax Deep Survey with VST}

   \subtitle{VII. Evolution and Structure of Late Type Galaxies inside the Virial Radius of the Fornax Cluster}

\author{ M. A. Raj\inst{\ref{inst1} , \ref{inst2}}  \and E. Iodice \inst{\ref{inst1}} \and N. R. Napolitano \inst{\ref{inst3}, \ref{inst1}} \and M. Spavone\inst{\ref{inst1}} \and  H-S. Su\inst{\ref{inst4}} \and R. F. Peletier \inst{\ref{inst5}} \and T. A. Davis\inst{\ref{inst6}} \and N. Zabel\inst{\ref{inst6}} \and M. Hilker\inst{\ref{inst7}} \and S. Mieske\inst{\ref{inst8}} \and  J. Falcon Barroso \inst{\ref{inst9}, \ref{inst10}} \and M. Cantiello \inst{\ref{inst11}} \and G. van de Ven  \inst{\ref{inst12}} \and A. E. Watkins \inst{\ref{inst4}}\and H. Salo \inst{\ref{inst4}}\and P. Schipani \inst{\ref{inst1}} \and M. Capaccioli \inst{\ref{inst2}}\and A. Venhola \inst{\ref{inst5}} }

% R. F. Peletier\inst{\ref{inst3}} \and et al. }
         %\fnmsep\thanks{Just to show the usage of the elements in the author field}

   \institute{INAF-Astronomical observatory of Capodimonte, via Moiariello 16, Naples, I-80131, Italy \\email:{mariaangela.raj@oacn.inaf.it}  \label{inst1}
   \and
   University of Naples ``Federico II'', C.U, Monte Santangelo, Via Cinthia, 80126, Naples, Italy  \label{inst2}
   \and
   School of Physics and Astronomy,  Sun Yat-sen University Zhuhai Campus, 2 Daxue Road,  Tangjia,  Zhuhai,  Guangdong 519082,  P.R. China  \label{inst3}
   \and
   Division of Astronomy, Department of Physics, University of Oulu, Oulu, Finland \label{inst4}
   \and
   Kapteyn Astronomical Institute, University of Groningen, PO Box 800, 9700 AV Groningen, The Netherlands \label{inst5}
   \and
   School of Physics and Astronomy, Cardiff University, Queen’s Building, The Parade, Cardiff, CF24 3AA, Wales, UK \label{inst6}
   \and
   European Southern Observatory, Karl-Schwarzschild-Strasse 2, D-85748 Garching bei Munchen, Germany \label{inst7}
   \and
   European Southern Observatory, Alonso de Cordova 3107, Vitacura, Santiago, Chile \label{inst8}
   \and
   Instituto de Astrof\`isica de Canarias, C Via Lactea s/n, 38200 La Laguna, Canary Islands, Spain \label{inst9}
   \and 
   Departamento de Astrof\`isica, Universidad de La Laguna, E-38200 La Laguna, Tenerife, Spain \label{inst10}
   \and 
   INAF-Astronomical Abruzzo Observatory, Via Maggini, 64100, Teramo, Italy \label{inst11}
   \and 
   Department of Astrophysics, University of Vienna, T\"urkenschanzstrasse, 17, 1180 Vienna, Austria \label{inst12}
   }

\date{Received..; Accepted..}

\abstract
{We present the study of a magnitude limited sample ($m_B \le $16.6 mag) of 13 late type galaxies (LTGs), observed inside the virial radius, $R_{vir}\sim$ 0.7 Mpc, of the Fornax cluster within the Fornax Deep Survey (FDS).}
{The main objective is to use surface brightness profiles and $g-i$ colour maps to obtain information on the internal structure of these galaxies and find signatures of the mechanisms that drive their evolution in high-density environment, which is inside the virial radius of the cluster.}
{By modelling galaxy isophotes, we extract the azimuthally averaged surface brightness profiles in four optical bands. We also derive $g-i$ colour profiles, and relevant structural parameters like total magnitude, and effective radius. For 10 of the galaxies in this sample, we observe a clear discontinuity in their typical exponential surface brightness profiles, derive their ``break radius'', and classify their disc-breaks into Type II (down-bending) or Type III (up-bending).}
{We find that Type-II galaxies have bluer average ($g-i$) colour  in their outer discs while Type-III galaxies are redder. The break radius increases with stellar mass and molecular gas mass while decreases with molecular gas-fractions. The inner and outer scale-lengths increase monotonically with absolute magnitude, as found in other works. For galaxies with CO(1-0) measurements, there is no detected cold gas beyond the break radius (within the uncertainties). In the context of morphological segregation of LTGs in clusters, we also find that, in Fornax, galaxies with morphological type 5$<$ T$\leq$9 ($\sim$60 \% of the sample) are located beyond the high-density, ETG-dominated regions, however there is no correlation between $T$ and the disc-break type. We do not find any correlation between the average ($g-i$) colours and cluster-centric distance, but the colour-magnitude relation holds true.}
{The main results of this work suggest that the disc breaks of LTGs inside the virial radius of the Fornax cluster seem to have arisen through a variety of mechanisms (e.g. ram-pressure stripping, tidal disruption), which is evident in their outer-disc colours and the absence of molecular gas beyond their break radius in some cases. This can result in a variety of stellar populations inside and outside the break radii.}

\titlerunning{LTGs inside the Virial Radius of the Fornax Cluster}
\authorrunning{M.A.Raj et al. }
\maketitle
%
%________________________________________________________________

\section{Introduction} \label{intro}
Spiral galaxies and irregular galaxies, which fall under the late type galaxies (hereafter, LTGs) morphology classification, have been primarily studied in the past, mostly concerning the formation of grand design structures e.g. spiral arms  \citep{Lin1964}.
These galaxies are rich in atomic and molecular gas and because of that, they are actively forming stars. 
Their stellar and (cold and diffuse) gaseous discs are known to be sensitive to the effect of the environment and their evolution (star formation history) is slower \citep{Boselli2006a} than that of early type galaxies (lenticular and elliptical galaxies; hereafter ETGs). This makes them interesting probes of a cluster environment \citep[e.g.][]{Blanton2009, Hwang2010}. 

Studying the effect of the environment (e.g. field vs cluster), where LTGs are located, is vital for unravelling the formation of their unique substructures \citep[e.g.][]{Boselli2006a}. Moreover, LTGs are ideal systems to study stellar population gradients (bulge and disc, inner and outer regions) which may have come about as a consequence of external processes \citep[e.g.][]{Bell2000,Gadotti2001}. 
The relative abundance of galaxies with respect to their morphological types has been shown to be related to the density of their local environment with a fraction of 80 \% of LTGs in the field, 60 \% in clusters, and $\sim$ 0\% in the core of rich clusters \citep{Dressler1980, Whitmore1993}. Their infall into a cluster environment makes their interstellar medium prone to hydrodynamic interactions with the hot intracluster gas \citep[e.g.][]{BinneyTremaine1987}, and their peculiar morphologies have been the topic of interest, especially concerning their structure formation and star formation regulation through a series of intracluster processes like ram pressure stripping \citep{Gunn1972}, star formation quenching \citep[e.g.][]{Abadi1999}, or galaxy-galaxy harassment \citep[e.g.][]{Moore96}.

Studies have shown that the first stage of evolution of LTGs, as they enter a dense environment, is characterised by the formation of a strong bar, and an open spiral pattern in the disc \citep{Moore1998, Mastropietro2005}. Further on, their spiral arms and rings can be  wiped out by tidal interactions, as shown by galaxy formation and evolution simulations \citep{Mastropietro2005}. Recently, though, numerical simulations \citep{Hwang2018} have shown that interactions with other galaxies, especially ETGs, in high-density cluster regions can also cause an LTG to lose its cold gas but still have more star formation activity during the collision phase (cold gas interacting with the hot gas of the cluster). Thus, the location of a galaxy in the cluster is a primary parameter to correlate with formation of structure in galaxies.
One of the several methods used is to investigate this evolution through the detailed study of the surface brightness profiles and their morphology classification based on their substructures (e.g., lopsided or warped disc, bars, peanut shaped bulges, spiral arms).  

Particular insight can be gained from quantitative studies of the light distribution of stellar discs down to the faintest regions, exploring the formation of their substructures and evolutionary paths \citep[e.g.][]{Courteau1996,Trujillo2002}. Light distributions of disc galaxies are characterised by an exponential decline \citep{deVaucouleurs1959}. However, this so-called decline in edge-on spiral galaxies does not continue to its last measured point, but rather truncates after a certain radial scale-length \citep{vanderkruit1979}. The origin of these truncations can be explained though several models suggesting that accretion of external material in the discs of galaxies can produce extended discs which can be interpreted as truncations \citep[e.g.][]{Larson1976}. \citet{dejong1996b} predicted that there is substantial age gradients across these discs, which are not observed yet.

Alternatively, a truncation orginates by quenching of star formation activity due to the fall of gas density below a threshold for local instability e.g. interaction with other galaxies, tidal effect of the environment, stochastic fluctuations due to internal dynamics \citep{FallEfstathiou1980, Schaye2004}.  This star-formation threshold is also associated with disc breaks in light profiles of galaxies \citep[e.g.][]{Martin2001}. 

Truncations should not be confused with disc breaks. Originally, the former was found to occur in edge-on galaxies at 4.2$\pm$0.5 scale-lengths \citep{kruit81a, kruit81b}, while the latter occurs around 2.5$\pm$0.6 scale-lengths \citep{PohlenTrujillo2006}. \citet{kregel2004} found that the surface brightness at the estimated truncation radius was 25.3$\pm$0.6 mag arcsec$^2$ ($r$-band). On the other hand, disc breaks were found in the surface brightness range 23-25 mag /arcsec$^2$  \citep{PohlenTrujillo2006}. In some cases, disc breaks and truncations can co-exist, but the mechanisms causing them are different \citep{Nacho2012}. Depending on the shape of the surface- brightness profiles, disc breaks are classified into (i) Type I-no break (ii) Type II-downbending break (iii) Type III-upbending break \citep[see][]{Erwin05,PohlenTrujillo2006,Erwin2008}. 

Different mechanisms have been proposed to explain the origin of disc breaks. Type II disc breaks occur as a consequence of star-formation threshold \citep{FallEfstathiou1980, Schaye2004}, while Type III disc breaks have been associated to minor/major mergers \citep[e.g.][]{Younger2007, Borlaff2014}. Furthermore, disc breaks are located at closer radial distances to the centre in face-on galaxies than in edge-on systems. Internal mechanisms owing to the formation of bars and rings in spiral galaxies can also cause disc breaks \citep{Mateos2013}. As such, there has been research on disc breaks in galaxies of different morphologies \citep[e.g.][]{PohlenTrujillo2006} as well as a function of the environment i.e. field vs cluster \citep{Pranger2017}.

Clusters are suitable for the study of discs in LTGs as they allow to map the dynamical processes of the galaxies in a dense environment. After Virgo, the Fornax cluster is the nearest and second most massive galaxy concentration within 20 Mpc, with a virial mass of M = $9 \times 10 ^{13} M_{\odot}$ \citep{Drinkwater2001}. Fornax is dynamically more evolved than Virgo, as most of its bright ($m_B < 15$ mag) cluster members \citep{Ferguson1989} are ETGs which are mainly located in its core \citep{ Grillmair1994, Jordan2007}. However, the mass assembly of the Fornax cluster is still ongoing \citep{Scharf2005} and the intra-cluster light (ICL) in its core  indicates interactions between cluster members \citep{Iodice2017b,Pota2018,Spiniello2018}. \citet{Iodice2017b} found that this ICL is the counterpart in the diffuse light of the overdensity in the blue intra-cluster globular clusters \citep{Dabrusco2016, Mik2018} and a fraction of the ICL population are low-mass dwarf galaxies \citep{Venhola2017}. The observed number density drop of low surface-brightness (LSB) galaxies below a cluster-centric distance r = 0.6 deg (180 kpc) by \citet{Venhola2017} is a proof of the effect of the dense environment on the evolution of galaxies. In the same regions, the high-velocity planetary nebulae (PNe, \citealt{Spiniello2018}) and globular clusters (GCs, \citealt{Pota2018}) show that this ICL component is unbound to galaxies, possibly after gravitational interactions, and is dynamically old. All of these findings indicate that Fornax is an evolved, yet active environment and a rich reservoir for the study of the evolution and the structure of the galaxies in a cluster environment, specifically, inside the virial radius of the Fornax cluster. 
The aim of this work is to study the structure of the LTGs inside the cluster and address the diverse evolutionary paths these systems can take after falling into a cluster. In particular, by taking advantage of the deep Fornax Deep Survey (FDS) data, we can map the light distribution down to unprecedented limits in the Fornax cluster. This allows us to detect any asymmetries in the outskirts of the discs as well as tidal tails or streams that would indicate possible interactions. By analysing their surface brightness distribution, we aim at identifying disc-breaks and finding any correlations with the location of galaxies inside the cluster.

In this study, we present the analysis of LTGs i.e. spiral galaxies selected from \citet{Ferguson1989}, which are brighter than $m_B \le $16.6 mag inside the virial radius of the Fornax cluster. We give a short summary of the FDS in Sect. \ref{FDS_data}. We describe the procedure and method to derive the radial profiles, colours, and structural parameters like effective radius and total magnitude  in Sect. \ref{ana_dp}. We give a detailed description of the algorithm we have developed to derive break radii in Sect. \ref{al_br}. A surface photometry analysis of each galaxy presented in this work is described in Appendix \ref{ltgs}, with their corresponding images in surface brightness, $g-i$ colour maps and profiles, with surface brightness profiles shown in Appendix B.  We summarise our results in Sect. \ref{resu2} and \ref{tr_gi}, and discuss the formation of structures caused by secular evolution and the effect the environment in Sect. \ref{dis}. Conclusions and future perspectives are given in Sect. \ref{conc}. The methodologies of some of this work are presented in Appendix C. 
%__________________________________________________________________

\begin{table*}[h]
\caption{Image quality of FDS fields.}              % title of Table
\label{tab_psf}      % is used to refer this table in the text
\centering                                      % used for centering table
\begin{tabular}{ccccccccc}          % centered columns (4 columns)

\hline\hline                        % inserts double horizontal lines
Field 
& \multicolumn{2}{c}{$u$ band}
& \multicolumn{2}{c}{$g$ band } 
& \multicolumn{2}{c}{$r$ band}
& \multicolumn{2}{c}{$i$ band}
  % table heading
\\[+0.1cm]
\hline 
 &\multicolumn{1}{c}{$FWHM$}
 &\multicolumn{1}{c}{depth} 
 &\multicolumn{1}{c}{$FWHM$}
 &\multicolumn{1}{c}{depth} 
 &\multicolumn{1}{c}{$FWHM$}
 &\multicolumn{1}{c}{depth} 
 &\multicolumn{1}{c}{$FWHM$}
 &\multicolumn{1}{c}{depth} 
 \\
 & [arcsec]
 & [$\frac{mag}{arcsec^2}$]
 & [arcsec]
 & [$\frac{mag}{arcsec^2}$]
 & [arcsec]
 & [$\frac{mag}{arcsec^2}$]
 & [arcsec]
 & [$\frac{mag}{arcsec^2}$]
\\[+0.1cm]
 (1)
  & \multicolumn{2}{c}{(2)}
  &\multicolumn{2}{c}{(3)}
  &\multicolumn{2}{c}{(4)}
  &\multicolumn{2}{c}{(5)}\\
 \hline\hline
F2&1.21&26.89&1.11&28.35& 0.90&27.78&0.79 & 26.75 \\
F4 &1.18&26.95&1.39&28.45& 1.19&27.76&0.70 & 27.01\\
F5 &1.33&27.33&1.15&28.54& 1.39&27.85&1.08 & 26.97\\
F6&1.11&27.43&0.84&28.47& 1.08 &27.73&1.21& 26.82\\
F7&1.04&27.3&0.83&28.56& 0.95&27.81&1.42 & 26.62\\
F12&1.15&27.44&0.83&28.49& 1.04&27.84&1.17& 26.79\\
F15&1.30&27.12&1.13&28.35&0.90&27.89&0.97 & 26.87\\
F16&1.31&27.27&1.26&28.43& 0.94&27.84&1.08 & 26.96\\
F17&1.27&27.1&1.11&28.29& 0.87&27.96&1.01 & 26.92\\

\hline
\end{tabular}
\tablefoot{{\it Col.1 -}FDS fields ;{\it Col.2-} average seeing, and surface brightness corresponding to 1$\sigma$ S/N per arcsec for $u$-band; {\it Col.3-Col.5-} same as Col. 2 for $g,r,$ and $i$ bands. The PSF FWHM and depth of each filter in all fields are adopted from \citet{Venhola2018}. } 
%{\it Col.6 \& Col. 7}; {\it Col.7 \& Col. 8} }
 \end{table*}
\begin{figure*}
   \centering
   \includegraphics[scale=0.6]{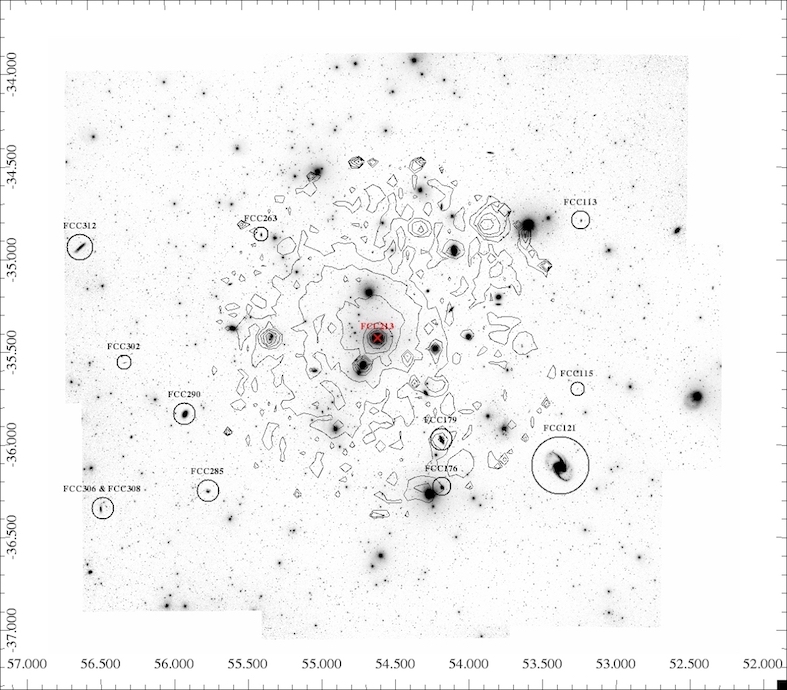}
      \caption{VST mosaic in the $g$-band of the Fornax cluster inside the virial radius, at about 9 sq deg.  The marked circles indicate the 12 LTGs presented in this work, with FCC267 located at the top north of the virial radius, outside this mosaic. Black
contours are the X-Ray emission from ROSAT \citep{Paolillo2002}. The X-ray contours are spaced by a factor of 1.3, with the lowest level
at 3.0 $\times$ 10$^{-3}$ counts/arcmin$^2$/s. }
           
         \label{cluster_g}
   \end{figure*}
\section{The Fornax Deep Survey: Data} \label{FDS_data}
The Fornax Deep Survey, observed with the ESO VLT Survey Telescope \citep{Schipani2012} is a deep, multi-band imaging ($u,g,r,i$) survey of the Fornax cluster, covering 26 square degrees around the core of the cluster \citep{Iodice2017}. 
This work focuses on the central 9 square degrees ($\alpha = 03h \ 38m \ 29.024s, \delta= -35d \ 27' \ 03.18'' $) which cover most of the cluster area inside the virial radius, $R_{vir} \sim 0.7$ Mpc \citep{Drinkwater2001}. 
FDS observations are part of the Guaranteed Time Observation surveys, FOCUS (P.I. R. Peletier) and VEGAS (P.I. E.Iodice, \citealt{Cap2015}) acquired with the ESO VLT Survey Telescope (VST) operating from Cerro Paranal. VST is a 2.6 m wide-field optical survey telescope with a wide-field camera, OmegaCam \citep{Kui2011}, that covers a 1x1 degree$^2$ field of view, and has a mean pixel scale of 0.21 arcsec /pixel. 
A thorough description of the observing procedure, data reduction pipelines, calibration, and quality-assessment of the final data products were presented by \citet{Venhola2017,Venhola2018} and \citet{Iodice2016, Iodice2017b}. The data presented in this work were obtained during several visitor mode runs in dark time for $u,g,r$ bands and in grey time for $i$- band (see \citealt[]{Iodice2018} for details on the observed FDS fields). Data were reduced with two pipelines: (i) VST-Tube \citep{Grado2012, Cap2015}; (ii) Astro-WISE \citep{Valentijn2007}. The VST-Tube was used to reduce data for the central 9 square degrees, while Astro-WISE was used for the data reduction of 26 square degrees. We use both data products in this work. 

The FDS fields were acquired using a \textit{step-dither} observing strategy, which consists of a cycle of short exposures of 150s each, centred on the core and its adjacent fields ($\le$ 1 deg) in the Fornax cluster. Fields with few or no bright objects were used to derive an average sky frame for each night and in all bands, which is scaled and subtracted from each science frame. This method has proven to provide an accurate estimation of the sky background around bright and extended galaxies \citep[see][]{Iodice2016, Iodice2017b, Iodice2018,Venhola2017, Venhola2018}. For each field, we obtained 76 exposures of 150s each in \textit{u} band, 54 in \textit{g} and \textit{r} bands, and 35 in \textit{i} band, thus resulting in a total exposure time of 3.17 hr in \textit{u} band, 2.25 hr in  \textit{g} and \textit{r} bands, and 1.46 hr in \textit{i} band. Images with seeing FWHM $\le $ 1.5 arcsec were used in the final production of co-added frames  (see \citealt{Iodice2018} for exposure times of each filter). The FWHM (arcsec) and depth ($mag/arsec^2$) of FDS fields in each filter ($u,g,r,i$) are given in Tab. \ref{tab_psf} \citep[see][]{Venhola2018}.

The $g$-band mosaic covering 9 square degrees around the core of the Fornax cluster, with marked circles indicating the galaxies studied and presented in this paper, is shown in Fig \ref{cluster_g}.\\
  This is a complete sample in $m_B \le $ 16.6 mag consisting of 13 LTGs inside the virial radius of the Fornax cluster (see Tab. 1). The sample of spiral galaxies has been selected from \citet{Ferguson1989}, with morphological type T $\ge$ 1. To be consistent with the catalog we refer to \citep[from][]{Ferguson1989} and keeping in mind that the morphological classification changes over time, we use a generic term, i.e., address the galaxies in our sample as LTGs. This work is complementary to that of  \citet{Iodice2018} on ETGs.\\
  A detailed description of each galaxy in the sample is given in Appendix A.

 \begin{table*}[h]
\caption{LTGs  brighter than $m_B \le 16.6$ mag inside the virial radius of the Fornax cluster}              % title of Table
\label{tab1}      % is used to refer this table in the text
\centering                                      % used for centering table
\begin{tabular}{cccccccc}          % centered columns (4 columns)
\hline\hline                        % inserts double horizontal lines
object & $\alpha$ & $\delta$  & Morph type & radial velocity
  & $m_B$ & FDS Field & Names \\    % table heading
            &h:m:s&d:m:s& &km/s&mag& & \\
  (1)&(2)&(3)&(4)&(5)&(6)&(7)& (8)\\
 \hline\hline
FCC113 & 03 33 06.8 & -34 48 29	&  ScdIII pec      &  1388   & 14.8   &    F15-16      &  ESO358-G15 \\
FCC115&03 33 09.2 & -35 43 07& Sdm(on edge)& 1700&16.6& F16& ESO358-G16\\
FCC121&03 33 36.4&-36 08 25&SB(s)b&1635&10.32&F16-F17&NGC1365, ESO358-G17\\
FCC176 & 03 36 45.0 & -36 15 22	&  SBa (SAB(s)a)  &  1414  & 13.74       &    F12-F17   & NGC1369, ESO358-G34 \\
FCC179 & 03 36 46.3 & -35 59 58	&  Sa	          &  868    & 12.09 &    F11-F12    &  NGC1386, ESO358-G35 \\ 
FCC263 & 03 41 32.6 & -34 53 17	&  SBcdIII           &  1724 &    14.04  &    F6             &  ESO358-G51\\
FCC267&03 41 45.4& -33 47 31&SmIV&834&16.1&F4-F5&\\
FCC285 & 03 43 02.2 & -36 16 24	&  SdIII	          &  886    &   14.1       &     F7            & NGC1437A, ESO358-G54\\
FCC290 & 03 43 37.1 & -35 51 13	&  ScII	          &  1387  &  12.41               & F6-F7 & NGC1437, ESO358-G58\\
FCC302&03 45 12.1& -35 34 15&Sdm (on edge)& 803& 15.5&F2-F6&ESO358-G060\\
FCC306 & 03 45 45.3 & -36 20 48	&  SBmIII	          &  886    &   15.6 & F7&  \\
FCC308 & 03 45 54.7 & -36 21 25	& Sd	          &  1487   & 13.97 &  F7 & NGC1437B, ESO358-G61\\
FCC312  & 03 46 18.9 & -34 56 37 & Scd	          &  1929   & 12.83 & F2-F6 & ESO358-G63\\
\hline
\end{tabular}
\tablefoot{{\it Col.1 -} Fornax cluster members from \citet{Ferguson1989}; {\it Col.2 \& Col.3 -} Right Ascension and Declination; {\it Col.4 - }
 Morphological type; {\it Col.5 -} Heliocentric radial velocity obtained from NED; {\it Col.6 -} total magnitude in $B$ band as given in NED\footnote{The NASA/IPAC Extragalactic Database (NED) is operated by the Jet Propulsion Laboratory, California Institute of Technology, under contract with the National Aeronautics and Space Administration.}; {\it Col.7 -} Location in the FDS Field; {\it Col.8 -}  Alternative catalog names} 
 \end{table*}

\section{Analysis: Surface Photometry} \label{ana_dp}
In this section, we give a brief description of the method we adopt, following \citet{Iodice2017b, Iodice2018}, to derive galaxy parameters:  total magnitude, effective radius and stellar mass to light ratio (M/L) ratio.

\subsection{Method: Isophote Fitting} \label{iso}We extracted azimuthally averaged intensity profiles for each object from the sky-subtracted images in four respective bands, using the \textsc{ellipse} \citep{ellipse87} task in IRAF \footnote{IRAF is distributed by the National Optical Astronomy Observatories, which are operated by the Association of Universities for Research in Astronomy, Inc., under cooperative agreement with the National Science Foundation} \citep[for a thorough explanation of this method, see][]{Iodice2018}. 

The main steps are as follows: 
\begin{enumerate}
\item Create masks for bright objects (galaxy and stars) around our galaxy of interest 
\item  Fit isophotes in elliptical annuli, starting from the centre of the galaxy, up to its outer edge in the FDS field (out to $\sim$ 0.5 deg). 
We keep the geometric centre of symmetry fixed while the ellipticity and position angle are free parameters. For lopsided galaxies (FCC113 and FCC285) and barred galaxies (FCC263 and FCC121), we use the galactic centre of one band as the same centre for other bands so that the same regions are mapped in the study and characterisation of their structures. 
\item From the intensity profiles, we (i) estimate the limiting radius corresponding to the outer most annulus $R_{lim}$ (see Tab. \ref{tab_lim}) \footnote{Note that $R_{lim}$ is different from the depth of the image.}, where the galaxy's light blends into the average background level \footnote{The average background level is the residual after subtracting the sky frame, this results in a value close to zero \citep[see][]{Iodice2016}.} and (ii) derive the residual sky background from the outer annuli of all galaxies in each band.
\end{enumerate}

\begin{table}[h]
\caption{Limiting radius of the intensity profiles}              % title of Table
\label{tab_lim}      % is used to refer this table in the text
\centering                                      % used for centering table
\begin{tabular}{ccc}          % centered columns (4 columns)
\hline\hline                        % inserts double horizontal lines
object & $R_{lim}$ & $\mu$ \\   
 & arcsec& mag/arcsec$^2$\\
  (1)&(2)&(3)\\
 \hline\hline
FCC113 &42.23 &25.7$\pm$ 0.19 \\
FCC115 &36.04 & 25.5 $\pm$0.11\\
FCC121 &436.7 & 26.6 $\pm$0.32\\
FCC176 & 86.4 & 26.5$\pm$ 0.4\\
FCC179 & 139.2 & 25.83 $\pm$0.3\\
FCC263 & 71.41 & 25.7$\pm$ 0.15\\ 
FCC267 & 42.23 & 26.82$\pm$ 0.4\\
FCC285 & 86.4 & 26.0$\pm$ 0.07\\
FCC290 & 139.2 & 26.44$\pm$ 0.03\\
FCC302 & 51.1 &26.94$\pm$ 0.21\\
FCC306 & 22.75 &25.8$\pm$ 0.1\\
FCC308 & 115.0 & 26.3$\pm$ 0.13\\
FCC312  & 153.1 & 25.13$\pm$ 0.14\\
\hline
\end{tabular}
\tablefoot{{\it Col.1 -} Fornax cluster members from \citet{Ferguson1989}; {\it Col.2-} Limiting radius in $r-$band ; {\it Col.3-} Surface brightness at the limiting radius. }
\end{table}

\subsection{Products: total magnitude, effective radius, colour, and stellar mass to light ratio} \label{dp}
We adopt the procedure of \citet{Iodice2018} to derive parameters: total magnitude, effective radius, colour, stellar mass to light ratio, for all galaxies in our sample. 
\begin{enumerate}
\item The resulting output of the isophote fit is used to provide their respective intensity profiles, from which we derive the azimuthally averaged surface brightness (SB) profiles. This is followed by a correction for the residual background level estimated at $R \ge R_{lim}$, in each band for all the galaxies in our sample.
The error estimates on magnitudes take into account the
uncertainties on the photometric calibration and sky subtraction \citep[see][]{Cap2015, Iodice2016, Iodice2018}. 

\item From the azimuthally averaged SB profiles, we derive $g-i$ colour profiles for each galaxy (Appendix B). We also show the $g-i$ colour maps for each of these galaxies (Appendix B), obtained from the images. 
\item From the isophote fits, we use a growth curve analysis to derive total magnitude and effective radius in each band for all galaxies (see Tab. \ref{tab2}). 
\item We derive average ($g-i$) and ($g-r$)  colours for each of these galaxies from the radial profiles (see Tab.  \ref{tab2}).
\item From the average ($g-i$) colour, we estimate the stellar mass $M_*$ by using the empirical relation $log_{10}$ $\frac{M_*}{M_{\odot}}$ = 1.15 + 0.70($g-i$)- 0.4$M_i$ from  \cite{Taylor2011}, where $M_i$  is the absolute magnitude in $i$- band \footnote{The empirical relation proposed by \cite{Taylor2011} assumed a Chabrier IMF.}. According to \cite{Taylor2011}, this relation provides an estimate of the stellar mass-to-light ratio $M_*/L_i$ with 1 $\sigma$ accuracy of $\sim$ 0.1 dex. The $M_*/L_i$ value for each galaxy of the sample is given in Tab. \ref{tab3}. 
\end{enumerate}

A full illustration of the surface brightness profiles, $g-i$ colour profiles and maps for individual galaxies are shown in Appendix B. 
\begin{table*}[h]
\caption{Derived Parameters of LTGs inside the virial radius}              % title of Table
\label{tab2}      % is used to refer this table in the text
\centering                                      % used for centering table
\begin{tabular}{lcccccccc}          % centered columns (4 columns)
\hline\hline                        % inserts double horizontal lines
object & $m_u$ & $m_g$ & $m_r$ & $m_i$ & $ Re_u $& $ Re_g$ & $ Re_r$ & $ Re_i$ \\   
 & mag& mag& mag& mag & arcsec& arcsec& arcsec& arcsec  \\
  (1)&(2)&(3)&(4)&(5)&(6)&(7)& (8)& (9)\\
 \hline\hline
FCC113 &$15.51  \pm    0.04$ &$15.01\pm  0.02$ &$14.44\pm  0.01$&$14.3\pm  0.01$ &$29.74  \pm   0.76$ &$19.51   \pm   0.19 $&$20.56   \pm   0.1 $&  $  20.64   \pm    0.108$\\
FCC115 & $ 17.18  \pm     0.01 $ &$16.35    \pm   0.01$ & $15.91  \pm     0.39$ &$15.94     \pm  0.01$ & $17.9   \pm    0.5  $&$19.54   \pm   0.11   $  & $ 20.58 \pm     0.11$&$      16.58 \pm   0.2$\\
FCC121 & $ 10.65 \pm     0.05 $ &$9.33   \pm   0.04$ & $8.79  \pm     0.03$ &$8.43     \pm  0.03$ & $134  \pm    3.98 $&$142.4  \pm   3.69   $  & $ 109.8 \pm     2.15$&$      132.8  \pm   2.89$\\
FCC176 & $ 13.95  \pm     0.01 $ &$12.51    \pm   0.01$ & $11.74   \pm     0.01$ &$11.45      \pm  0.01$ & $56.76   \pm    1.59  $&$53.03   \pm   0.69   $  & $ 53.73  \pm     2.28$&$       47.42    \pm   1.20$\\
FCC179 &$13.17 \pm   0.02$ & $11.44  \pm   0.01 $& $ 10.68\pm  0.01 $ & $10.34    \pm  0.01$ &$22.75\pm0.42$ &$ 29.68 \pm 0.50 $& $30.03 \pm0.50$ & $28.01\pm  0.73$\\
FCC263 & $14.72     \pm   0.05$ &$12.91     \pm   0.05$ & $12.7    \pm   0.02$ & $12.58     \pm   0.01$ &$21.82     \pm  1.57 $&$      25.36    \pm   1.67   $ &$  27.15   \pm   0.56 $&$     21.53    \pm  0.14 $\\ 
FCC267 & $ 16.6  \pm     0.08 $ &$15.68    \pm   0.03$ & $15.09  \pm     0.01$ &$14.8    \pm  0.01$ & $14.1   \pm    0.5  $&$17.6   \pm   0.4  $  & $ 19.4  \pm  0.4 $&$       20.0   \pm   0.3$\\

FCC285 &$14.57    \pm  0.03$ &$13.15    \pm  0.02 $ & $ 13.02   \pm     0.03$ & $ 12.77   \pm    0.01$& $40.84  \pm1.63$&$46.78 \pm 1.29$&$ 49.9   \pm 1.54 $&$  51.1   \pm   0.64$\\
FCC290 &$12.6   \pm  0.095 $&$ 11.4  \pm 0.01  $&$      11.08  \pm 0.01 $&$    10.74  \pm  0.01 $&$ 49.68  \pm     0.53  $&$       49.65   \pm   0.58     $&$  48.52  \pm    0.16  $&$     48.41  \pm   0.06$\\
FCC302 & $ 16.29  \pm     0.09 $ &$15.58   \pm   0.01$ & $15.37  \pm     0.04$ &$15.06  \pm  0.1$ & $28.49   \pm    1.7 $&$21.92 \pm   0.75$  & $ 24.15\pm     0.78$&$      28.84  \pm  1.64$\\
FCC306 & $16.54   \pm   0.02$& $15.29    \pm    0.03$&$ 15.18    \pm   0.01$ & $ 15.0    \pm  0.014$& $ 9.01\pm   0.16   $&$   10.52    \pm  0.15  $&$     9.7   \pm  0.06  $&$     9.85  \pm     0.06$\\
FCC308 & $14.36    \pm  0.02 $ & $12.87   \pm   0.01$& $ 12.54  \pm     0.01$ & $ 12.17     \pm   0.02 $&$ 53.65 \pm       1.41 $&$      44.91   \pm   0.56    $&$   37.11   \pm   0.82 $&$      46.46     \pm 0.37$\\
FCC312  &$13.38     \pm  0.05$ &$ 11.4      \pm 0.02 $ & $ 10.89     \pm   0.09$&$ 10.45 \pm    0.04$ & $49.34  \pm    2.13   $&$   76.0 \pm      3.67  $&$     109.5    \pm   9.33   $&$    117.8  \pm     7.26$\\
\hline
\end{tabular}
\tablefoot{{\it Col.1 -} Fornax cluster members from \citet{Ferguson1989}; {\it Col.2 to Col.5 -} Total magnitude in the $u$, $g$, $r$ and $i$ bands respectively, derived from the isophote fit. Values were corrected for the galactic extinction, using the absorption coefficient by \citet{Schlegel98}; {\it Col.6 to Col.9 -} Effective radius in the $u$, $g$, $r$ and $i$ bands respectively, derived from the isophote fit.}
\end{table*}

 \begin{table*}[h]
\caption{Stellar mass estimates of LTGs in the $i$ -band}              % title of Table
\label{tab3}      % is used to refer this table in the text
\centering                                      % used for centering table
\begin{tabular}{lccccccc}          % centered columns (4 columns)
\hline\hline                        % inserts double horizontal lines
object & $D_{core}$ & $M_i$ & $R_e$ & $g-r$ & $ g-i $& $ M_*$ & $ M/L$ \\   
 & deg& mag& kpc& mag & mag& $10^{10} M_{\odot}$&  \\
  (1)&(2)&(3)&(4)&(5)&(6)&(7)& (8)\\
 \hline\hline
FCC113 & 1.21&-16.74 &1.98&$0.57\pm  0.03$ &$0.71  \pm   0.02$ &0.02&0.56\\
FCC115 &1.05&-15.78 &1.59 &$0.43    \pm  0.02$ & $0.41  \pm    0.02  $&0.01& 0.35\\
FCC121 &1.06&-22.58&12.74 &$0.54     \pm  0.02$ & $0.90  \pm    0.02 $&6.49 & 0.77\\
FCC176 & 0.82&-19.85 & 4.55&$0.77      \pm  0.02$ & $1.06   \pm     0.02  $&0.68 & 0.99\\
FCC179 &0.55 & -20.7& 2.68& $0.76   \pm  0.02$ &$1.10 \pm 0.02 $ &1.58& 1.06 \\
FCC263 & 0.79&-18.08& 2.06& $0.21  \pm   0.07$ &$0.32     \pm   0.06  $&0.04&0.3\\ 
FCC267 & 1.73& -15.24 &1.92&$0.59   \pm  0.04$ & $0.88  \pm     0.04  $&0.01&0.74\\

FCC285 &1.17 &-17.43&4.90 & $ 0.13  \pm    0.05$& $0.38   \pm 0.04$&0.02&0.33\\
FCC290 &1.05&-20.49 &4.64&$   0.33  \pm  0.04 $&$ 0.66  \pm     0.04  $&0.64&0.52\\
FCC302 & 1.30&-15.92& 2.76 &$0.21 \pm  0.02$ & $0.52  \pm     0.02 $&0.01& 0.41\\
FCC306 & 1.69& -15.21&0.94& $ 0.11  \pm  0.04$& $ 0.29 \pm    0.04   $&0.003&0.29\\
FCC308 & 1.69& -17.49&4.46& $ 0.33     \pm   0.02 $&$ 0.69  \pm     0.03 $&0.04&0.55\\
FCC312  &1.59 &-20.89 & 11.30&$ 0.51 \pm    0.11$ & $0.95  \pm     0.06   $&1.48&0.83\\
\hline
\end{tabular}
\tablefoot{{\it Col.1 -} Fornax cluster members from \citet{Ferguson1989}; {\it Col.2-} Projected distance from the galaxy centre in degree, i.e. from NGC 1399 (FCC213); {\it Col.3-} Absolute magnitude in $r$-band, derived using the distance modulus from NED and \citet{Tully2009}; {\it Col.4 -} Effective radius (kpc) in $i$-band;{\it Col.5 \& 6-} Average $g-r$ and $g-i$ colours ;   {\it Col.7 \& Col. 8-} Stellar mass and mass-to-Light (M/L) in $i$-band. }
\end{table*}

\section{Analysis: the break radius} \label{al_br}
One of the main novelties of this work is the algorithm we have developed to derive break radii of galaxies. It is based on concepts used in literature to derive break radii \citep[e.g.][]{PohlenTrujillo2006}, but it is fully automatised and reproducible, as it is based on coded rules to define the break radius and bootstraps the results over a variety of initial conditions for the best-fitting procedure. 

In this section, we illustrate the main steps to derive the break radius from surface brightness profiles, and provide a detailed description 
of the algorithm  we use. 

\subsection{Surface brightness profile deconvolution} \label{SB-dec}
We derive the break radius from surface brightness profiles in the $r$-band as this is one of the highest transmission OmegaCam filters 
and least-affected by dust absorption, which could be quite high in LTGs.

Since the break radius is derived on the disc regions of LTGs, it is important to account for the effect of the contamination from their bright cores, which scatters their light in the regions around them because of the effect of the point spread function (PSF). The mathematical effect introduced by the PSF over the measured 2D light distribution is equivalent to a convolution of a matrix of the intrinsic intensity values (the surface brightness) and a 2D kernel (the matrix corresponding to the PSF). In order to correct this, we need to invert this operation i.e., perform a deconvolution, assuming the kernel is given.

The first step of deconvolution (assuming the kernel is given) is to accurately measure the PSF out to a radial distance comparable to that of the galaxies' discs. 
\citet{Cap2015} found that the scattered light can considerably affect the local light profiles in VST images of galaxies of various sizes in different ways. 
To fully account for the broadening effect of the seeing on their sample of galaxy images, they characterized the PSF from the VST images. 

\begin{figure*}
%\vspace{-2 cm}
 \centering
  \includegraphics[width=\hsize]{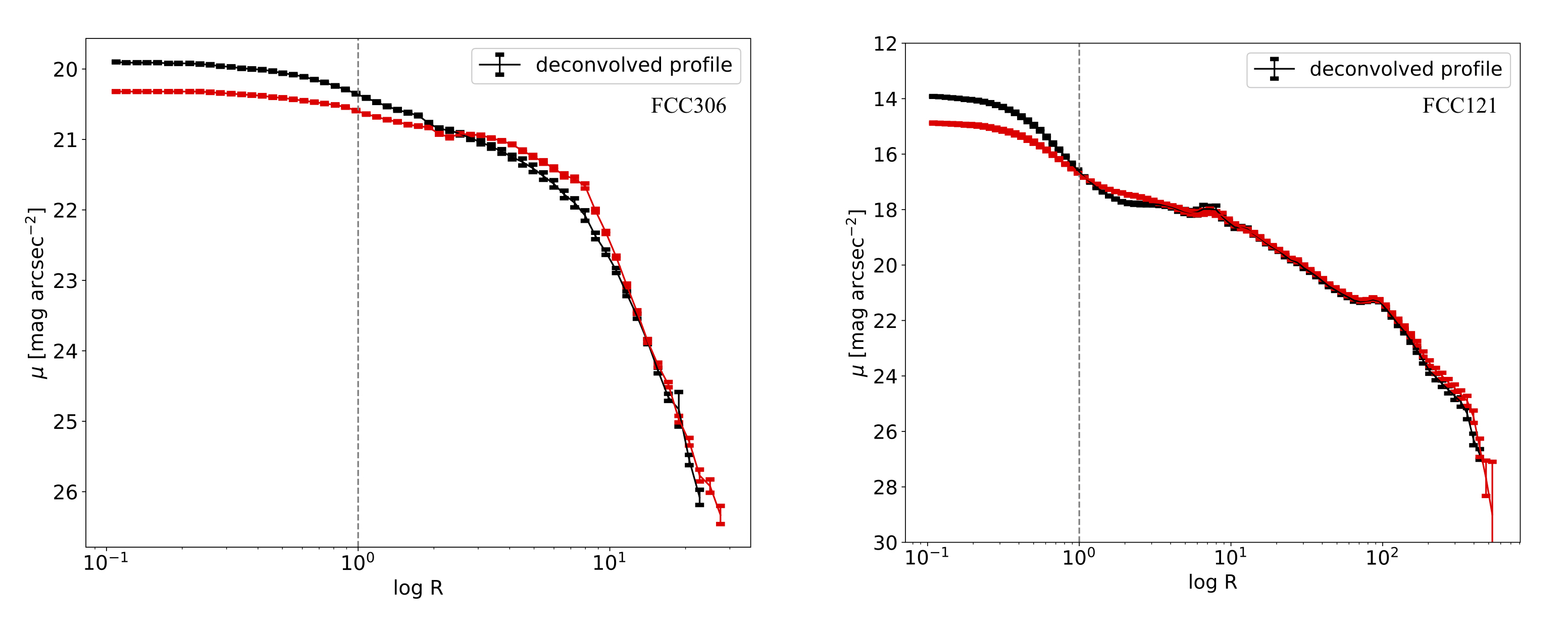}
      %\vspace{-3.5cm}

   \caption{The deconvolved SB profiles over-plotted on the original SB profiles (red) for the smallest, FCC306 ($R_e $ =0.94 kpc in $i$-band) and the largest galaxy, FCC121 ($R_e $ =12.74 kpc in $i$-band) in the left and right panels. The PSF (1 arcsec) for FDS data is marked in grey dotted lines.  }
   \label{psf_decon}
\end{figure*}

The sample presented in this work consists of galaxies of different angular extent (see Fig. \ref{psf_decon}.
FCC306 is the smallest galaxy in our sample (R$_e$ =0.94 Kpc in $i$-band) and FCC121 is the largest galaxy,  (R$_e$ =12.74 Kpc in $i$-band). From Fig. \ref{psf_decon}, it is clear that the effect of the PSF (1 arcsec) scatters the light from the central regions to the outer regions. 
The effect mitigates outside the typical size of the PSF, where, for the conservation of the flux, the light scattered from the core makes the observed profile brighter than the intrinsic (deconvolved) profile. Finally, the effect of the PSF is completely canceled far away from the center, typically outside 10 arcsec. 
Thus, in order to minimise the effect of the PSF on our structural analysis, we deconvolve galaxies with the PSF, by using the Lucy-Richardson algorithm \citep{Lucy74,Richardson72}, as done by  \citet[hereafter SP+17]{Spavone2017}. A full description of the deconvolution method used for VST data is presented by Spavone et al. (in prep). 
 The robustness of the deconvolution algorithm is demonstrated in Appendix~\ref{test_dec}, where we compare the performance against the method from \citet{Borlaff17}.

The surface brightness profiles and break radii are derived from the deconvolved images in $r$- band. We also derive the deconvolved profiles in $g$ and $i$ bands. The deconvolved profiles ($r$-band) are given in  Fig.\ref{truncs} and Fig.\ref{Atruncs}, and deconvolved ($g-i$) colour profiles are shown in Appendix. B.  %\ref{algo}.

\subsection{Algorithm: Break radius} \label{algo}
 Since discs of LTGs show exponential SB profiles, which become linear in the log-linear plane, the break radius can be defined as the radius where the logarithmic SB, defined as $\mu$ =-2.5 log $I(R)$, where $I(R)$ is the SB as a function of the radius $R$, shows a discontinuity in the slope. The presence of this discontinuity is first guessed by fitting the disc with a single linear model using the following equation,
\begin{equation}\label{eq:fit}
     \mu(R)=\beta + \alpha R
 \end{equation}
 where $\alpha$ is the slope of the linear fit and $\beta$ is the intercept. The best fit is performed via least square linear fitting of 1D surface brightness profile in $r-$band (deconvolved). The least square approximation of the linear system is based on the Euclidean norm of the squared residuals. 
 
 This is followed by evaluating the residuals of the best fit. If the residuals show monotonic deviations from the best fit (rms residuals $>$ 0.5 mag)\footnote{The residuals in irregular galaxies correspond to the presence of wiggles in their profiles. We check this before concluding that it is associated with disc-breaks.} that increase with radius, it suggests that the profile has a break and needs a second component. 
Then, two linear fits are implemented, using Eq. \ref{eq:fit}. In order to take into account different intervals in radius i.e., before and after the discontinuity, we have developed a procedure which automatically defines these intervals (radial ranges) to determine the break radius ($B_r$). 
The linear fits are performed outside the bulge component. For most galaxies, the region where the bulge dominates was taken from \citet{Salo2015}. Though we initially consider the whole profile (including inner components), we exclude the breaks detected in the inner regions as they are connected to component-transition rather than disc-breaks \citep[e.g.][]{Laine2014, Laine2016}.

Since the aim of this multi-linear fit approach is to study the disc regions of LTGs and detect any break in this component, it cannot be considered as a tool to decompose the observed profiles into different sub-components. In fact, it makes use of a simple linear regression to define the regions where there is a significant slope change\footnote{This also accounts for a slope error $>$ 0.05.} in the light profiles of LTGs to measure the transition point.
 
 In more detail, the first step is to estimate the radial range where the disc component is defined (i.e., an inner disc limit, $range_{in}$, and an outer disc limit, $range_{out}$). This is done by an initial guess of the disc scale-lengths ($h_{in}$ and $h_{out}$), from which the range is derived by varying `$n$' \footnote{Here $n$ is defined as the number of SB points obtained from \textsc{ellipse}, using a semi-major axis (pixels) step of 0.1 (1.0 + $step$)} number of data points ($max_{range_{in}}$ = $range_{in} \pm n$;  $max_{range_{out}}$ = $range_{out} \pm n$, we used n=2). The procedure is repeated for the second sub-component and so on (in case there is more than one evident break radius, known from the residuals of the fit).
 
We stress here that the determination of the fitting range only extends to the limiting radius $R_{lim}$. \\
Once the domains of the different disc components are defined, a linear least square fit is done to each of these ranges ($range_{in}$ and $range_{out}$), producing $(n+1)^2$ best fits\footnote{This combination is obtained by \textit{rule of product} i.e., $n+1$ ranges for the upper and lower limits of the radial scale-lengths } for the inner and outer ranges. The $\alpha$ and $\beta$ parameters (in Eq.\ref{eq:fit}) of each of these best linear regression fits are stored and used to estimate the point of intersection, otherwise called the break radius ($B_r$).

\begin{figure*}
   \centering
   \includegraphics[width=17cm]{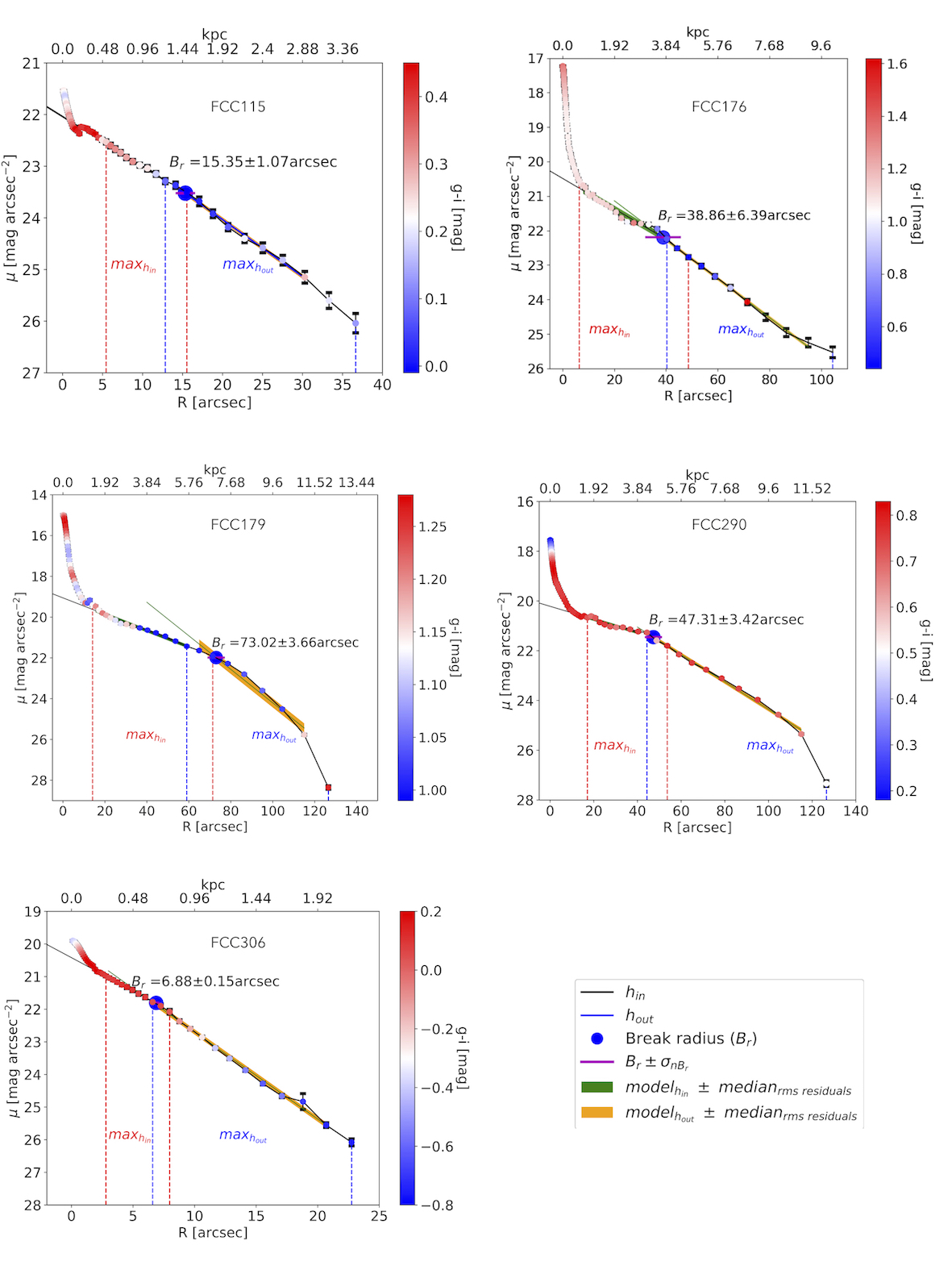}
      \caption{Surface brightness profiles of galaxies with Type II disc break, with ($g-i$) colour as a third parameter. In each plot, the break radius ($B_r$) is marked at the intersecting point of the linear fits performed between $h_{in}$ and $h_{out}$, with $\sigma_{nB_r}$ as the median of $(n+1)^3$ combinations of best fits on the inner and outer scale-lengths. The vertical dashed lines ($max_{h_{in}}$ and $max_{h_{out}}$) indicate the regions for $max_{range_{in}} = range_{in} \pm n $ and $max_{range_{out}} =  range_{out} \pm n$ where the algorithm produces $(n+1)^2$ linear least square fits. The shaded regions on $h_{in}$ and $h_{out}$ indicate the median of the  rms of the residuals for $(n+1)^2$ linear least square fits. }
             \label{truncs}
   \end{figure*}

\begin{figure*}
  \centering
   \includegraphics[width=17cm]{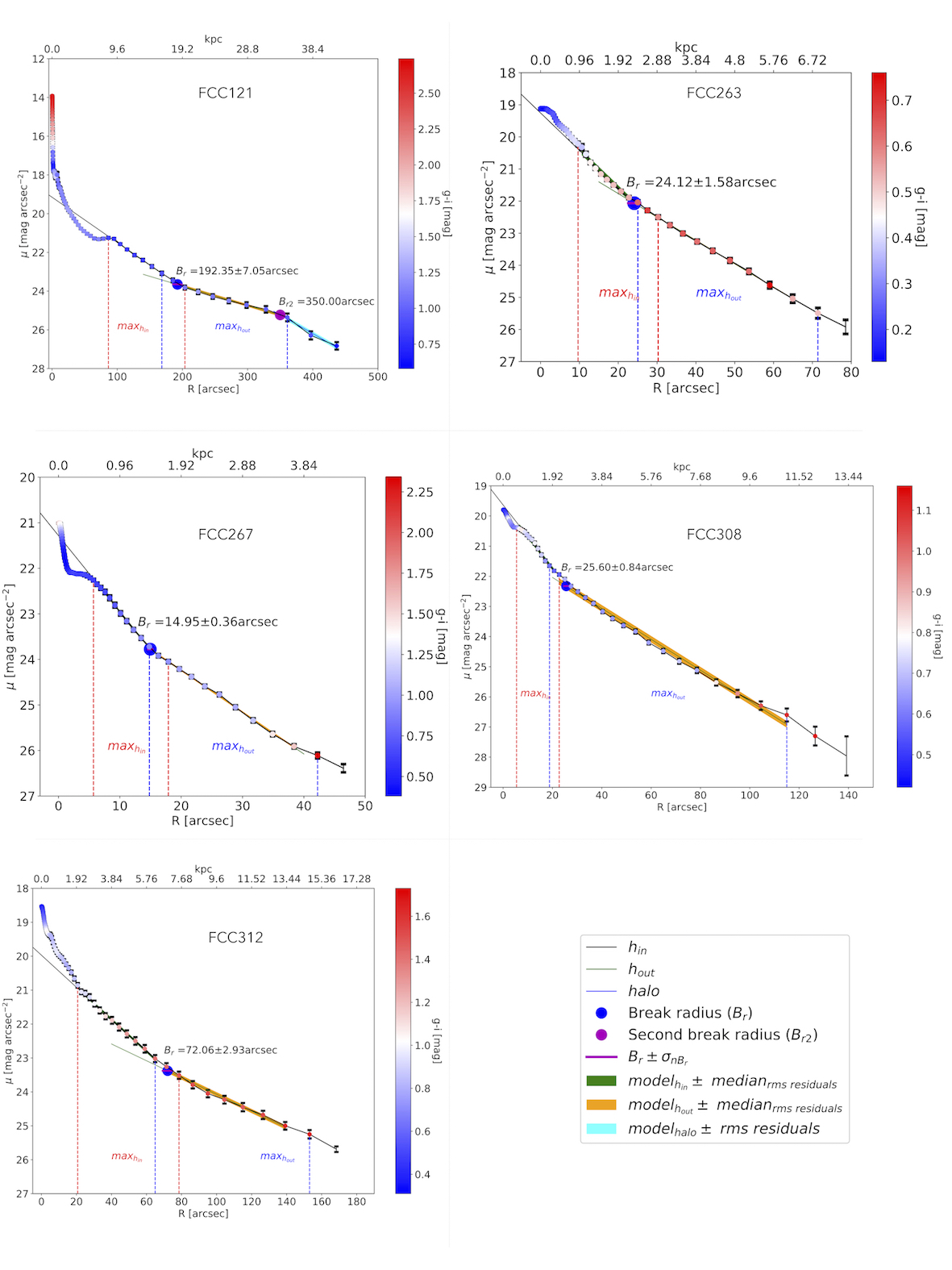}
      \caption{Surface brightness profiles of galaxies with Type III disc break, similar to Fig. \ref{truncs}.}
           \vspace{0.35cm}
           
         \label{Atruncs}
   \end{figure*}

The combinations of the fitting for the ranges $range_{in}$ and $range_{out}$ are chosen such that they do not coincide and all points of the disc regions on the surface brightness profile are considered. The selection of these combinations is different from the selection of $max_{range_{in}}$ and $max_{range_{out}}$where these represent the total range considered for $range_{in}$ and $range_{out}$, but not for the fitting itself. This in turn produces $(n+1)^3$ estimates of the break radius. The median of $(n+1)^3$ intersecting points is chosen as the final break radius. The rms residuals (the numerical rank of the scaled Vandermonde matrix) for each linear regression fit are computed and the median of this is shown as the range of variance of the fitted models' scale-lengths $h_{in}$ and $h_{out}$ (see Fig.\ref{truncs} and Fig. \ref{Atruncs}). To show the regions of variance of the intersecting point from the $(n+1)^3$ best fits, the standard deviation of $(n+1)^3$ break radii ($\sigma_{nBr}$) is marked as a significant error on the estimation of the break radius. This procedure is repeated by varying the initial guess of the disc scale-lengths until a minimal standard deviation of the $(n+1)^3$ break radii is obtained, resulting in the best selection for $h_{in}$ and $h_{out}$.

The regions (here,  $range_{in}$ and $range_{out}$ obtained from the best fit iteration) are given as the disc scale-lengths $h_{in}$ and $h_{out}$ and are used to derive the average ($g-i$) colour of $h_{in}$ and $h_{out}$.

The break radius (in arcsec and kpc) of 10 galaxies with their $\sigma_{nBr}$, surface brightness at the break $\mu_{B_r}$, and average ($g-i$) colour (derived from deconvolved profiles) for the inner $h_{in}$ and outer $h_{out}$ discs are listed in Tab. \ref{tab4}, with their deconvolved surface brightness profiles (in $r$- band), shown in Fig. \ref{truncs} and Fig. \ref{Atruncs}. 
We also apply the algorithm on irregular galaxies: FCC113, FCC285, and FCC302. In such cases, we do not detect disc-breaks, but rather wiggles and bumps associated to the irregularities (star forming regions) present in these galaxies. This is also checked by subtracting the small-scale wiggles from the profiles of these galaxies, and applying the algorithm on the obtained profile. These three galaxies show a Type-I profile. 

Most of the galaxies in the sample show clear evidence of a single break radius, however we cannot exclude that further substructures are present beyond their outer disc scale-lengths. 

For example, FCC121 shows a second break (see Fig. \ref{Atruncs}) at $\mu(R)>25.5$ mag arcsec$^{-2}$, which is well detected by the algorithm but it is not discussed in our analysis. This second break could be associated with the truncation radius that occurs in the outermost optical extent of a galaxy \citep[see e.g.][]{Nacho2012}.
Also, we need a minimum of 3 data points to produce $(n+1)^2$ fits (with a lower limit of $n$=1), that is within the $R_{lim}$. For this reason, we do not derive a second break for some galaxies of the sample (e.g. FCC267, FCC308) though it seems to be present in their profiles.

An example of the $(n+1)^3$ best fits with minimal standard deviation, produced by the algorithm is shown in Appendix \ref{C}. Parameters of the best fit for LTGs with disc breaks are given in Tab. \ref{tab5}.

For all galaxies, the break radii (inner and outer disc-breaks as in the case of FCC121) are located within regions where the effects of PSF-convolution are negligible. This is proven by deriving the break radius from deconvolved profiles and the results (original vs deconvolved profiles) are the same, as shown in  Appendix. \ref{dec_orig} and \ref{dec_orig2}.

\raggedbottom
\begin{table*}[h]
\caption{Parameters of LTGs with a disc break}              % title of Table
\label{tab4}      % is used to refer this table in the text
\centering                                      % used for centering table
\begin{tabular}{ccccccccccc}          % centered columns (4 columns)

\hline\hline                        % inserts double horizontal lines
object 
& \multicolumn{2}{c}{$B_r $}
&$\sigma_{nB_{r}}$   
&$\mu_{B_{r}}$     
& \multicolumn{2}{c}{$  h_{in} $ } 
& \multicolumn{2}{c}{$ h_{out}$ }
& Hubble
& Break    % table heading
\\[+0.1cm]
 &\multicolumn{1}{c}{[arcsec]}
 &\multicolumn{1}{c}{[kpc]} 
 &  [arcsec]                              
 & [$\frac{mag}{arcsec^2}$]
 &\multicolumn{1}{c}{[arcsec]}
 & \multicolumn{1}{c}{$g-i$ [mag]}
 & \multicolumn{1}{c}{[arcsec]}
 &\multicolumn{1}{c}{$g-i$[mag]}
 & Type (T)
 & Type
\\[+0.1cm]
  (1)
  & \multicolumn{2}{c}{(2)}
  &\multicolumn{1}{c}{(3)}
  &\multicolumn{1}{c}{(4)}
  &\multicolumn{2}{c}{(5)} 
  &\multicolumn{2}{c}{(6)}
  &\multicolumn{1}{c}{(7)}
  &\multicolumn{1}{c}{(8)}\\
 \hline\hline
FCC113 &  - &  - & - &  -  & - & -  & -  & - &  6 &I\\
FCC115   &   15.35 &  1.47 & 1.07  &  23.52   & 6.84 & 0.25 $\pm$ 0.03  &   16.15 & 0.12 $\pm$ 0.03    & 8  &II \\

FCC121  &   192.35  & 18.46 & 7.05   &   23.65   &63.9 &  0.88 $\pm$ 0.02   &  175.7 &  1.09 $\pm$ 0.04   &  3 &III+II\\

FCC176  &   38.86    & 3.73   & 6.39  &  22.20  & 28.66 &  1.10 $\pm$ 0.02  &  54.73  & 0.76 $\pm$ 0.2  & 1&  II \\

FCC179  &   73.02  &  7.01&3.66&   21.98   & 43.47 & 1.10 $\pm$ 0.02 &   50.08& 1.05 $\pm$ 0.02 &1 & II \\

FCC263  &   24.12  &  2.32&1.58&   22.06   &14.42 & 0.49 $\pm$ 0.02 &  37.39&0.63 $\pm$ 0.02 & 2& III \\

FCC267  &   14.95  &  1.43 &0.39  &   23.83    & 8.52 & 0.76 $\pm$ 0.03   &  22.11 &  1.26 $\pm$ 0.13 &9&III\\ 
FCC285&  - &  - & - &  -  & - & -  & -  & -   &  7 &I\\
FCC290  &   47.31  &  4.54&3.42 &  21.45 & 23.22  & 0.72 $\pm$ 0.03&  70.66& 0.71$\pm$ 0.02 &4 &II\\
FCC302 &  - &  - & - &  -  & - & -  & -  & -  &  8 &I\\
FCC306  &   6.88  &  0.66&0.15  &   21.81  & 2.92 & 0.09 $\pm$ 0.01&  14.1& -0.39 $\pm$ 0.07 &9& II \\
FCC308  &   25.60 & 2.46 &0.84 &   22.32   & 14.11 & 0.71 $\pm$ 0.02 &  92.25 &0.75 $\pm$ 0.04 & 7& III \\
FCC312  &72.06  & 6.91&2.93 &   23.38  & 42.17&      1.16  $\pm$ 0.05  & 67.79 & 1.54 $\pm$ 0.03 &7 & III\\

\hline
\end{tabular}
\tablefoot{{\it Col.1 -} LTGs with a disc break; {\it Col.2 -} Break Radius in units of arcsec and kpc (1 arcsec = 0.096 kpc);  {\it Col.3 -} Standard deviation of the break radii from $(n+1)^3$ combinations; {\it Col.4- } Surface brightness at the break radius; {\it Col.5 -} inner scale-length in units of arcsec, and average $g-i$ colour for $h_{in}$; {\it Col.6-}  outer scale-length in units of arcsec, and average $g-i$ colour for $h_{out}$; {\it Col.7-} Hubble Type T;{\it Col.8-} Profile classification \footnote{FCC267 and FCC308 are also classified as Type III+II, but we do not mention them in Tab. \ref{tab4} due to lack of data points to be fitted with the algorithm (see Sect. \ref{algo}).}} 
 \end{table*}

\section{Morphological segregation of LTGs inside the virial radius of the Fornax cluster} \label{resu2}
\begin{figure}[h]
%\hspace{-0.55cm}
% \vspace{-0.25cm}
   \includegraphics[width=9cm]{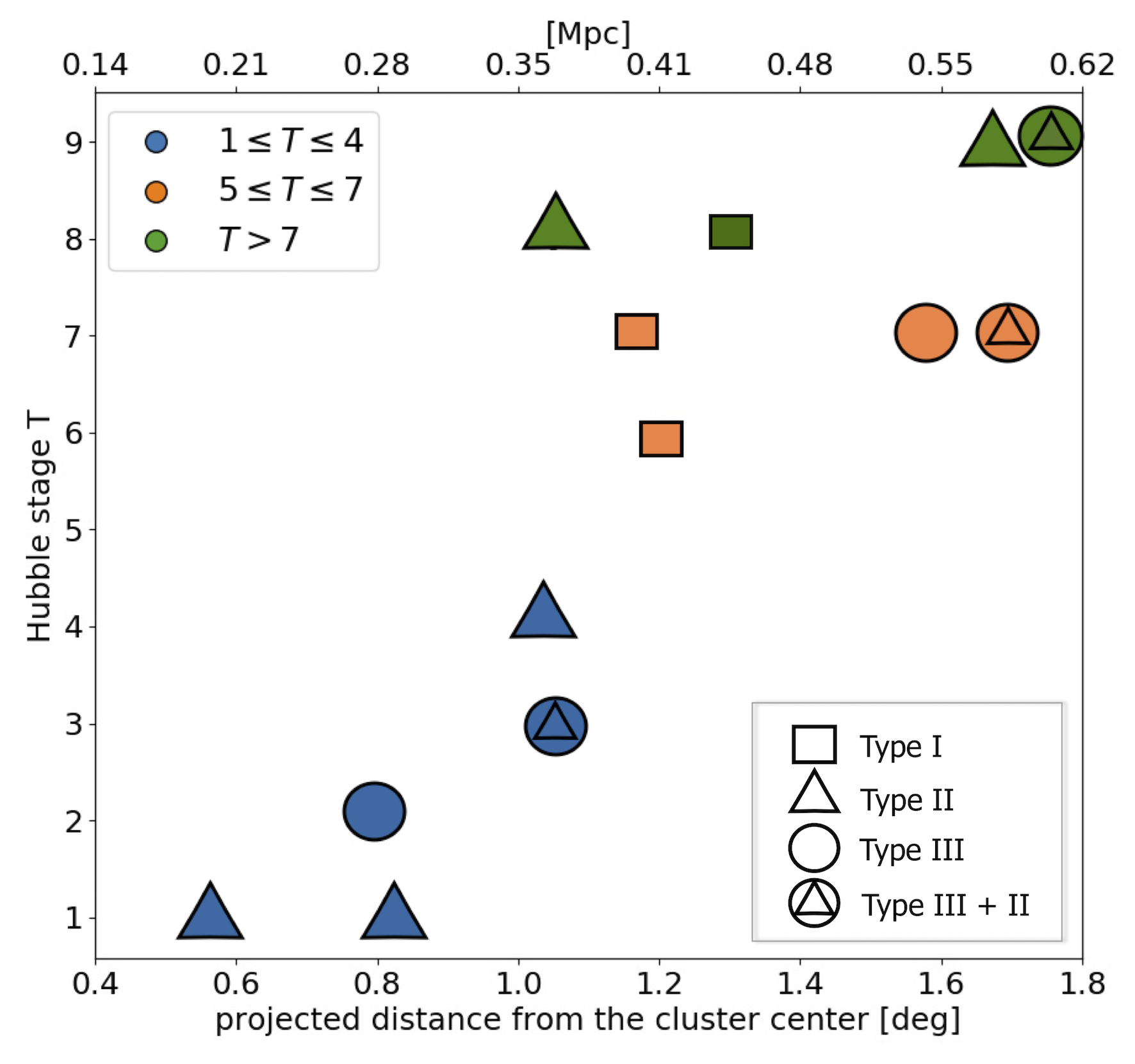}
      \caption{Hubble Type T as a function of projected cluster-centric distance. For each galaxy, their disc-break type is represented as squares (Type I) or triangles (Type II), or circles (Type III). For 3 galaxies with secondary breaks other than their primary break-type, they are represented as circle+triangle(Type III+II). }
    \label{T_dist}
   \end{figure}

The sample of LTGs inside the virial radius of the Fornax cluster is heterogeneous in morphology as measured by their T-type Hubble classes. 
These are shown as a function of the projected distance in Fig. \ref{T_dist}. The correlation is not as tight as previously reported in other similar environments e.g., Virgo \citep{Bing87} and other clusters \citep{Whitmore1993}. Here, the projected distances have been used as a primary parameter to discern 
the formation and evolution of the galaxies' substructures as a function of the cluster environment. However, one must also take into account other environment-independent mechanisms like 
pre-processing \citep[e.g.][]{Yutaka2004}, which are independent of the local environment. %and could have contributed to their morphology at the position where they are currently observed. 
In this section, we analyse the morphological structures of the galaxies in our sample by grouping them into bins of Hubble stage $T$ as a function of projected cluster-centric distance ($D_{core}$), shown in Fig. \ref{T_dist}.

The LTGs in Fornax with morphological type  $1 \le T \leq 4$ are FCC121, FCC176, FCC179, FCC263, and FCC290. 
 These galaxies are located within $D_{core} \leq$ 1.1 deg from the cluster centre (see Fig. \ref{cluster_g} and \ref{T_dist}), 
 corresponding to 0.54 $R_{vir}$. Overall, they are among the most luminous ($m_B <$ 14 mag, refer Tab. \ref{tab1}) LTGs in our sample. 
 They show regular spiral or barred-spiral structures with clear grand design features. 
Three of the above mentioned galaxies (FCC176, FCC179, FCC263) are, in projection, located within the X-ray halo of NGC 1399 (see Fig. \ref{cluster_g}). 

FCC176 and FCC179 are redder than other LTGs inside the virial radius (see Fig. \ref{col_dist}) with the former devoid of atomic and molecular gas \citep{Fuller2014, Schroder2001} and the latter shows dust and molecular gas in its spiral arms.
FCC121 and FCC290 are located beyond the X-ray halo with morphological type $T > 3$. 
FCC290 has spiral arms only inside 1$R_e$ (see Appendix \ref {FCC290}). FCC121 has grand design spiral structure with a bar.

Galaxies with morphological type $5 \le T \leq 7$ are FCC113, FCC285, FCC308, and FCC312. These galaxies are located beyond $D_{core} >$  1.1 deg.
The former two galaxies (FCC113 and FCC285) are lopsided with luminosities 14 $<m_B <$ 15 mag and the latter two (FCC308, FCC312) have boxy discs with $m_B <$ 14 mag.
FCC113 and FCC285 exhibit irregular star-forming regions, which are evident in their SB images (Appendix. \ref{FCC113} and \ref{FCC285}). 
They are located on either sides of the cluster centre with their asymmetric discs elongated towards the cluster centre (see Fig. \ref{cluster_g}). 
FCC308 and FCC312 have irregular star-forming regions in their ill-defined spiral arms. Their thick discs also have flares which are signs of minor mergers  \citep[e.g][]{Karen2012}.  

The very late morphological type $ T > 7$ galaxies are FCC115, FCC302, FCC306, FCC267. 
FCC115 and FCC302 are edge-on galaxies (making it hard to determine their true  morphological T-type), while FCC306 is a bright dwarf galaxy \citep{Drinkwater2001}. 
FCC267 has a double nucleus or a dust-obscured bar (see Appendix \ref{FCC267}). 
These galaxies have irregular star-forming regions and lumosities in the range 15 $<m_B \leq$ 16.6 mag (refer Tab. \ref{tab1}). 
All the galaxies in this group are located beyond $D_{core} >$ 1 deg. 
None of the aforementioned ($ T > 7$) have visible, regular spiral structures with star-forming regions, unlike the galaxies at smaller cluster-centric distances.

\begin{figure}[h]
 \hspace{-0.25cm}
  \vspace{-0.25cm}
   \includegraphics[width= 10 cm]{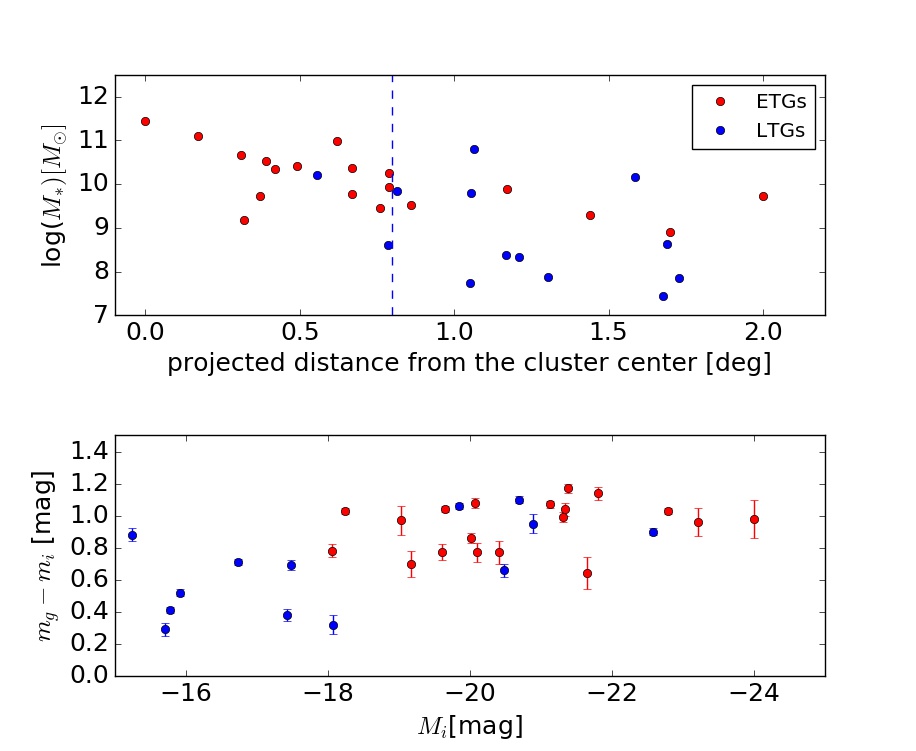}
   \caption{Stellar mass as function of cluster-centric distance (top panel) and colour-magnitude relation (lower panel) for LTGs and ETGs \citep[data from][]{Iodice2018} inside the virial radius of the Fornax cluster. }
    \label{gi_mass}
   \end{figure}

\begin{figure}
%\hspace{-0.2cm}
%\vspace{-1.5cm}
 %\centering
   \includegraphics[width=9.9cm]{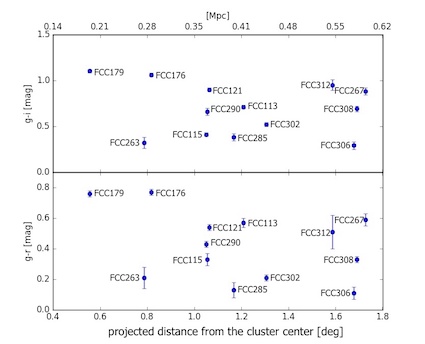}
 %  \vspace{-3.5cm}
      \caption{$g-i$ (top panel) and  $g-r$ (lower panel) colours as a function of projected distance from the cluster centre in degrees and Mpc. }
         \label{col_dist}
   \end{figure}

Despite our small heterogeneous sample of LTGs, we can still infer that the overall morphological segregation of galaxies inside the virial radius strongly 
suggests that the number density of ETGs is higher near the cluster centre ($D_{core}$ $< $0.8 deg $\sim$ 0.27 Mpc, \citealt{Iodice2018}), where only the massive LTGs of earlier types ($1 \leq T \leq 4$) are found (see Fig.~\ref{gi_mass}: top panel). 
Galaxies with morphological type 5 $<T \leq$ 9 ($\sim$ 60 \% of the sample) are located beyond $D_{core} >$  1 deg $\sim$ 0.36 Mpc i.e. beyond the high-density (ETG-dominated) regions defined by  \citet{Iodice2018}.

We do not find any trends between average ($g-i$ and $g-r$) colours  as a function of cluster-centric distance (see Fig. \ref{col_dist}), as observed by \citet{Iodice2018} for ETGs.
From the colour-magnitude relation diagram (Fig. \ref{gi_mass}), we find a positive correlation as expected and a clear segregation in colours from the bright ETGs in the cluster (see Fig.~\ref{gi_mass}: lower panel).

\section{Analysis of the disc breaks} \label{tr_gi}

In this section, we analyse disc breaks by investigating possible correlations with some global properties such as colours, total magnitudes, stellar mass,
molecular gas, and with the galaxy's location in the cluster. 
Since the sample is limited to 13 galaxies, the results cannot be considered on a statistical basis. However, it represents the complete sample of bright LTGs in the Fornax core and the analysis of their disc properties will provide important information on the assembly history of the cluster.

In Fig.~\ref{T_dist}, we show the different disc-break types are distributed as a function of the cluster-centric distance.
There is no evident correlation between the morphological type (T) and the disc-break type.
On the whole, we can give the same conclusion about the distribution of different disc-break types inside the virial radius, i.e. Type II and Type III breaks are found in galaxies at both small and large cluster-centric distances.
It is worth noting that Type I galaxies (only 3 galaxies) are found around $\sim$ 1.2 deg ($\sim0.4$~Mpc).

\begin{figure}
%\hspace{-0.5cm}
%\vspace{-1.5cm}

   \includegraphics[width=9.9cm]{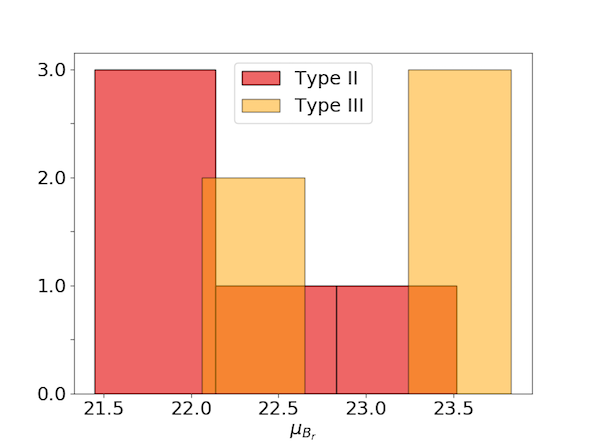}
   %\vspace{-3.5cm}
      \caption{Histogram of surface brightness at the break radius ($\mu_{B_r}$) for Type II and Type III galaxies.}
         \label{mu_br}
   \end{figure}

The surface brightness at the break radius in Type II discs is in the range 22.5 $\le$ $\mu_{B_r}$ $\le$ 24 (see Fig \ref {mu_br}). 

\begin{figure*}
 \centering
   \includegraphics[width=15cm]{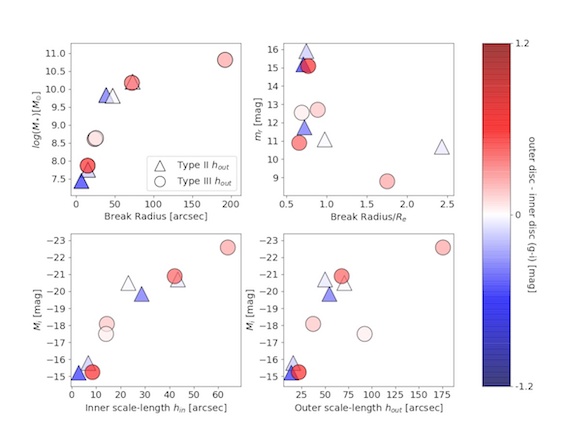}
      \caption{Analysis of galaxies with a break in their surface brightness profiles. Triangles represent Type II profiles, and circles represent Type-III, with average ($g-i$) colour of $h_{out}$-$h_{in}$ as colour map. Top panel (left):Stellar mass as a function of break radius; Top panel (right): Total magnitude as a function of break radius normalised to effective radius; Bottom Panel: Absolute magnitude in $i$-band as a function of  inner scale-length $h_{in}$(left) and outer scale-length $h_{out}$ (right).}
             
         \label{br analysis}
   \end{figure*}

On average, these galaxies have bluer  ($g-i$) colours in their outer discs ($h_{out}$) in comparison to their inner discs ($h_{in}$) (see Fig \ref{br analysis} and refer Tab. \ref{tab4}). FCC179 and FCC290 have bluer outer discs (than their inner discs), with a difference $\leq 0.05$ mag, which is above the error (refer Tab. \ref{tab4}). However, their inner discs have dust and this can cause the observed red colour. 
FCC115 and FCC306 are very late type galaxies that are bluer on average (see Fig. \ref{col_dist}) with the former being the faintest in the sample and the latter, a bright dwarf. 

Type III disc-breaks have surface brightness levels in the range  21.5 $\le$ $\mu_{B_r}$ $\le$ 23 (see Fig. \ref{mu_br}), i.e., more luminous than the $\mu_{B_r}$ of Type II disc-breaks.
 
These galaxies have redder ($g-i$) outer discs than their inner discs (see Fig. \ref{br analysis} and Tab. \ref{tab4}). 
Most of the Type III galaxies show signs of merger events taken place in the past e.g., FCC263 with disturbed molecular gas \citep{Zabel2018}, tidal tails \footnote{Galaxies with tidal tails were classified as Type IIIa by \citet{Watkins2019}} in FCC308 and FCC312 , double nuclei in FCC267, which can produce up-bending profiles. 
The redder colour may be accounted for the transfer of stellar mass to the outer disc during interactions \citep{Younger2007}. 
The redder outer discs may also be associated with the exhaustion of gas via ram-pressure stripping \citep[see e.g.][]{Steinhauser2016, Pranger2017}.

We find a positive correlation between stellar mass and the break radius (see Fig.\ref{br analysis}: top left panel). This trend is expected and has been shown before in literature  \citep[e.g][]{PohlenTrujillo2006}. In comparison to the results by \citet{PohlenTrujillo2006}, we also find that luminous galaxies have larger inner disc scale-lengths as shown in Fig. \ref{br analysis} (bottom panel). 
This trend is still evident in the plot of the total magnitude as a function of the break radius normalised to the effective radius (see Fig. \ref{br analysis}: top right panel), where most of our sample galaxies have break radius in the range of 0.5 $R_e < B_r$ $<$1.25 $R_e$, while the two brightest
galaxies have  $B_r \geq 1.7 R_e$.

\begin{figure}[h]
 \hspace{-0.25cm}
  \vspace{-0.25cm}
   \includegraphics[width=9.5 cm]{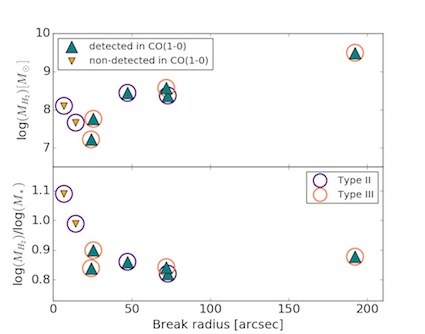}
   \caption{Molecular gas mass (top panel) and molecular gas-fractions (lower panel) as a function of break radius. Green triangles represent galaxies detected in CO(1-0), and orange triangles represent galaxies non-detected in CO(1-0) for eight LTGs in our sample, which were surveyed by \citet{Zabel2018}. Orange circles are galaxies with Type II profile, and purple circles are galaxies with Type III profile.} %Left panel: Stellar mass vs molecular gas mass \citep{Zabel2018}; Right Panel: Molecular gas mass as a function of break radius.}
    \label{co}
   \end{figure}

We used the molecular gas mass $M_{H_2}$ derived from CO (1-0) emission, as part of the ALMA Fornax cluster Survey \citep{Zabel2018} to obtain molecular gas-fractions ($M_{H_2}/M_{\ast}$) and plot them against the break radius for eight galaxies in our sample, of which six are detected in CO(1-0), and two are non-detected (see Fig. \ref{co}). 
The $M_{H_2}$ in the latter cases is the 3$\sigma$ upper limit. We also plot the CO maps in contours on our $g$-band images along with the break radius (see Appendix B). From this, we find that the molecular gas (CO) detection is within the uncertainties of the break radius. 
Since molecular  gas is a prerequisite for star formation, the break radius could also be a sign of a `break in star formation' 
\citep[see][]{Roskar2008, Sanchez2009,Christlein2010}. Fig.~\ref{co} shows that galaxies with larger break radius have higher molecular gas mass, and lower gas-fractions. 

\section{Summary of the results and Discussion}\label{dis}

Inside the virial radius of the Fornax cluster ($\leq 0.7$~Mpc), there are 13 LTGs brighter than $m_B \leq16.6$~mag. %The fainter galaxies are dwarfs.
In this work, we investigate the structure of this small, yet complete sample of 13 LTGs and in doing so, analyse any possible correlations with their position in the cluster.
One aim is to study the disc component and hence, we have developed an algorithm to define and measure break radii in their surface brightness profiles (Sect. \ref{tr_gi}).
  
The main results of this work are summarised below:
\begin{enumerate}

\item From Fig. \ref{T_dist}, it is evident that our sample of LTGs is heterogeneous in morphology. Still we find that galaxies with morphological type T $\leq$ 4 located at smaller cluster-centric distances. Among these, FCC176 and FCC179 (morphological type T=1) seem to be in a transition phase to S0/SB0. 
Later morphological types (5 $<T\leq$ 9, $\sim$ 60 \% of the sample)  are located at $D_{core}  >$ 1 deg $\sim$ 0.36 Mpc i.e. beyond the high-density of ETG-dominated regions ($D_{core}$$>$ 0.8 deg $\sim$ 0.27 Mpc). Most of these show signatures of tidal interactions in the form of lopsidedness (FCC113 and FCC285), thick discs and tidal tails (FCC308 and FCC312) or double nuclei/dust-obscured bar (FCC267).

    \item There is an equal fraction (38\%) of Type II and Type III discs in the sample, while only three galaxies (23\%) show a classical Type I exponential disc. There is no evident correlation between $T$ and the type of disc-break. Type II galaxies have bluer outer discs($h_{out}-h_{in} =-0.37$ mag), while Type III galaxies have redder outer discs($h_{out}-h_{in} =0.25$ mag). Brighter and massive galaxies have larger break radii. 
    For galaxies detected in CO(1-0), the molecular gas is within the break radius.

\end{enumerate}

In the following sections, we discuss the results mentioned above in order to address the evolution of LTGs inside the virial radius of the Fornax cluster.

\subsection{Evolution of LTGs inside the virial radius of the Fornax cluster} \label{morph}

The Fornax cluster has been considered as a classic example of a virialised cluster with regular galaxy distribution \citep{Grillmair1994, Jordan2007}. This cluster is dynamically more evolved than the Virgo cluster with a high fraction of  ETGs in its core  \citep{Grillmair1994, Jordan2007, Iodice2018}. Inside its virial radius, \citet{Iodice2018} found that the ETGs are not spherically distributed around NGC 1399, rather, they are located along a stripe i.e. N-S direction in the West side of NGC1399 (also refer to Fig. 1). Along this direction, signatures of an ICL component have been found \citep{Iodice2017}, mirrored by globular clusters (GC) density substructures  \citep{Dabrusco2016, Mik2018}, and supported by kinematical measurements by PNe \citep{Spiniello2018} and GCs \citep{Pota2018}. All of this suggests that there are interactions between Fornax cluster members \citep{Iodice2017} and among galaxies with the overall cluster potential (\citealt{Spiniello2018, Pota2018}). This ``stripe'' along the W-NW direction of NGC1399, otherwise called as W-NW sub-clump, may have resulted from the accretion of a galaxy-group during the build-up of the cluster, hence creating what has been perceived to be an asymmetry  \citep{Iodice2018}. To further substantiate this accretion and the consequences on its surroundings, all the LTGs brighter than $m_B\leq$16.6 mag, presented in this work, are not located spatially anywhere near this sub-clump of ETGs. In fact, these LTGs are located in the opposite direction (East side of NGC 1399, in 2D projection) of this high-density ETG-zone (Fig. 1), except for FCC113, a lopsided galaxy. However the lopsided tail of FCC113 appears to be elongated in the direction pointing towards the cluster centre (see Appendix. \ref{FCC113}). 

Our results concerning the morphological segregation of ETGs crowding the central regions \citep{Iodice2018}, and LTGs located at larger cluster-centric distance inside the virial radius of the Fornax cluster, is consistent with previous findings of the this relation in Virgo \citep{Bing87} and other clusters \citep{Dressler1980,Whitmore1993}. Some of the mechanisms that have been proposed to explain the morphological segregation are ram pressure stripping \citep{Gunn1972}, galaxy harassment \citep{Moore96}, truncated star-formation \citep{LarTin80}, galaxy-galaxy interaction \citep{Lavery88}. 

Due to the higher intra-cluster gas density and higher cluster-centric velocities of galaxies, ram-pressure stripping is 16 times more effective/stronger in the Virgo cluster \citep{Davies2013}. However, as the Fornax cluster hosts $\sim$ 300 galaxies with $B_T \sim$ 18  \citep{Ferguson1989}, which is an order of magnitude lower than the galaxy population in Virgo, it is suggested that galaxy-galaxy interactions play a prominent role in the evolution of Fornax cluster members \citep[e.g.][]{Dabrusco2016, Iodice2017, Iodice2017b, Venhola2017, Venhola2018, Spiniello2018}. 
In addition to this, it is also necessary to take into account pre-processing of a galaxy before falling into a cluster, and that the enhanced ram-pressure stripping occurs when the group passes through the cluster pericentre \citep{Vijay2013}. 

With these mechanisms stated, we examine the morphological evolution of LTGs inside the virial radius of the Fornax cluster. 

\subsubsection{Late-type spiral (Scd-Sdm) galaxies}
Galaxies with morphological type 5 $<T\leq$ 9 are located at $D_{core} > $ 1 deg $\sim$ 0.36 Mpc. Among these, FCC308 and FCC312 show signs of interactions in the form of tidal tails. The presence of these tails can either be explained by minor-merging events or by the disruption of the outer discs during infall, due to the strong tidal shear in the cluster centre \citep{Whitmore1993}. Galaxies with asymmetric stellar discs (FCC113 and FCC285) could have experienced similar mechanisms during infall, but after the cluster collapse \citep{Whitmore1993}. Since these galaxies have tidal disturbances similar to NGC 1427A, another possible mechanism for this is that it was triggered by a recent fly-by of another galaxy in the cluster \citep{Lee-Waddell2018}. However, more analysis on the HI  distribution of these galaxies are required to confirm the possibility of ram-pressure stripping acting on them \citep[e.g][]{Vollmer03}. 

FCC115, FCC302, FCC267, FCC306 are faint galaxies ($m_B \leq$ 16.6) with morphological type $T >$ 7 (Sd-Sdm). The former two have ill-defined spiral arms, and the latter two have faint spiral arms. As mentioned in the previous paragraph, these galaxies could have experienced disruptions due to the gravitational potential well of the cluster core, during infall. Since protogalactic clouds which are needed to form Sd galaxies are destroyed during the cluster collapse, these galaxies are dominant at larger cluster-centric distances \citep{Whitmore1993}. 

\subsubsection{Galaxies transitioning into S0}
FCC176 and FCC179, which are located at $D_{core}$ $\le$ 0.82 deg, are classic examples of galaxies transitioning into lenticulars (S0/SB0) in a dense environment. FCC290 ($D_{core}$ = 1.05) also has similar disc structure to FCC179, such that the spiral arms are found only in their central regions ($<$ 1$R_e$) whereas their outer-disc is devoid of any feature (including star-formation blobs and molecular gas). Their discs resemble the smooth structure, that is typically found in the discs of S0 galaxies. FCC176 and FCC179 are close to the cluster core and are the redder and more massive LTGs of our sample, with 
average $g-i$ colour $>$ 1 mag and $M_*/ 10^{10} M_{\odot} > $ 0.6 (see Tab.~\ref{tab5}). 
\citet{LarTin80} first explained how blue galaxies evolve to red galaxies over Hubble time, and this was followed by several other supporting evidences of galaxies losing their gas as a consequence of environmental effects \citep[e.g.][and references therein]{Boselli2006a, onofrio15, gao2018}. 
FCC176 could have formed before the cluster collapse and therefore resides in the X-ray halo of NGC 1399 \citep[e.g.][]{Whitmore1993}. This galaxy's morphological evolution under the influence of the cluster environment has caused it to be HI  deficient. However, FCC179 and FCC290 are detected in HI  \citep{Schroder2001} and CO(1-0) \citep{ Zabel2018}. The CO disc in spiral galaxies is usually concentrated within 1$R_e$ \citep[see e.g.][]{Davis2013}. The outer gas discs in spirals that extend to their optical radius are typically HI  dominated. HI  imaging would be required to confirm that these galaxies have been pre-processed \citep[see e.g.][]{Yutaka2004, Vijay2013}, or that they have experienced ram-pressure/tidal stripping during infall, which has resulted in the loss of their outer disc gas \citep{LarTin80}.

\subsection{Break radius as a proxy for the structural evolution of LTGs}

Disc breaks in the light profiles of LTGs have known to occur as a consequence of internal mechanisms (e.g. formation of bars) or external mechanisms (e.g. effect of the environment), or sometimes both \citep{Nacho2012}. It has proven to be a vital parameter in the study of the stellar and gaseous discs of LTGs \citep[e.g.][and references therein]{Martin2001,Roediger2012, Peters2017}. As such, we will use the break radius to further discuss the structural and morphological evolution of LTGs inside the virial radius of the Fornax cluster. 
To this aim, we also take into account the correlation of molecular gas distribution with the break radius, observed for six LTGs in the sample. The CO(1-0) detection \citep{Zabel2018} is within the uncertainties of the inner break radius (see Appendix B) implying that there could also be a star-formation break e.g. in FCC290, as previously suggested \citep[e.g.][]{Roskar2008, Sanchez2009, Christlein2010}.  As the concentration of molecular gas within the break radius could have occurred as a consequence of a different mechanism from that of the disc-break itself, further analysis on their stellar populations is required to confirm this correlation.

\subsubsection{Type III disc break}

Type~III breaks in clusters are found in galaxies with past minor/major mergers. Minor mergers can cause gas-inflows towards the centre of the primary gas-rich galaxy which steepens the inner profile, and expands the outer profile as the angular momentum is transferred outwards during the interaction \citep{Younger2007}, while major mergers usually produce Type-III S0 galaxies \citep{Borlaff2014}. 
This process can also explain the presence of molecular gas close the center, inside their break radius.

FCC308 and FCC312 have boxy discs and morphological type $T$ = 7, with flares in their disc, showing signs of minor merger, as already stated. The disc break of these galaxies could have also occurred as result of external mechanisms in the cluster environment, where the tidal shear of the cluster core causes disruptions in their disc. 

FCC121 and FCC263 are barred-spiral galaxies which do not have cold gas beyond their break radius. Within their break radius, these galaxies also show strong star-formation activity (molecular gas and bright blue knots, see Appendix. B), whereas the outer disc of FCC263 does not show any signs of current star-formation, and in the case of FCC121, there is a small fraction of bright knots along the spiral arms.  External mechanisms as previously stated, result in the removal of gas during infall via tidal stripping by the halo potential \citep{LarTin80}. 

\subsubsection{Type II disc break}

Models for Type~II breaks are linked to star-formation threshold \citep[e.g.][]{Martin2001} and radial migration of stars \citep[e.g.][]{Martinez2009, Roediger2012}. The stellar disc of early-type spirals tends to be more massive than their gaseous disc, which implies a star-formation threshold around their break radius \citep{Martin2001}. 

The disc break in barred galaxies has also been associated with the bar radius and the Outer Lindlblad Resonance(OLR) \citep[see][]{Mateos2013, Laine2014}, like in the case of FCC176.

Flared discs in highly inclined galaxies like FCC115 (faintest galaxy) and FCC306 (bright dwarf) can produce a less-pronounced break \citep[see e.g.][]{Borlaff16}. The disruptions in the outskirts of their discs may have resulted as a consequence of the gravitational potential well of the cluster core.

FCC179 and FCC290 (Type II) are spiral galaxies without cold gas in their outer disc (see Sect. \ref{resu2} and \ref{morph}).
The former is located within the X-ray halo (see Fig. \ref{cluster_g}), while the latter is located beyond the X-ray halo. The break radius of these galaxies are not only associated with the absence of molecular gas in their outer disc but also the absence of spiral arms. 

Though \citet{Erwin2012} found that there was an absence of Type~II profiles in Virgo cluster S0s, suggesting that the Type~II profiles can transition into Type~I, in a cluster environment, or that Type~I were turning into Type~II profiles (no models have predicted this yet), there might be other mechanisms that could also cause them to form a Type-III break. As such, the processes (internal and external) causing disc-breaks are a part of the evolutionary phases of LTGs in the Fornax cluster.

\subsubsection{What can the average ($g-i$) colour of the inner and outer discs of LTGs indicate?}
We find that Type II galaxies have bluer outer discs, while Type III 
galaxies have redder outer discs. In addition, the molecular gas detection 
is only within the break radius of six LTGs. Both findings might suggest 
that there  is a stellar population gradient across the disc \citep[e.g.][]{Bell2000}.
Since we do not find any segregation inside the virial radius of the cluster for Type~II and Type~III discs, this might indicate that different processes acted in different regions of the cluster. 

There have been numerous studies concerning the outer-colour of LTGs with disc breaks \citep[see e.g.][and references therein]{Bakos2008,Roediger2012, Laine2016,Watkins2019}, but no conclusive results have been given to explain these findings. 
External mechanisms like ram-pressure stripping (removal of cold gas) and  strangulation (removal of hot gas), or stellar migration \citep[e.g][]{Roskar2008} have been found to cause reddening in cluster LTGs \citep[e.g.][]{Steinhauser2016, Pranger2017}. 
The ram-pressure stripping or strangulation could have been responsible 
for the reddening in the outer Type~III discs in FCC263 and FCC121, 
which are located (projected distance) in the transition region from high-to-low density region of the cluster, where the X-ray emission is decreasing (see Fig.~\ref{cluster_g}).
For the other two Type~III galaxies (FCC267 and FCC312) far away from the cluster core, with a significant color gradient between inner and outer disc, a different process could have been responsible for the redder colors in the outskirts. Both of them, show also evident signs of past merging (see Appendix~\ref{FCC267} and Appendix~\ref{FCC312}).

Simulations by \citet{Hwang2018} have shown that LTGs entering a cluster could have encounters with ETGs, and during this phase, they can have strong star-formation activity yet losing their cold gas, which can also cause the presence of bluer outer discs. This could be the case of Type~II galaxies inside the high density regions of the cluster, as FCC176 and FCC179 that show a color difference larger than the error estimate (see Tab.~\ref{tab4}) and an ongoing star formation in the centre (see Appendix~\ref{FCC176} and Appendix~\ref{FCC179}).

\section{Concluding remarks and future perspectives} \label{conc}
FDS data allow us to map the light distribution of galaxies down to the faintest magnitudes where the effects of the environment on the evolutionary stages of Fornax cluster galaxies can be studied in depth. 
In this work we have shown how such studies are especially useful in the analysis of disc-breaks, which in turn provides a ground in further analysis of the stellar populations beyond the break radius. 

Despite the limited size of our LTG sample (13 objects), the morphological segregation of LTGs inside the virial radius of the Fornax cluster is clearly detected and is consistent with previous results that suggest that high-density ETG-dominated zones inhibit the formation of LTGs (5$ <$ $T\leq$ 9) in a cluster environment. 

We have used the results derived from the break radius as a means to discern the different mechanisms (external and internal) causing the structural evolution of LTGs inside the virial radius of the Fornax cluster. Their structural evolution would eventually cause a morphological transition. Since the sample is small, we do not give a generic conclusion that disc-breaks are indeed good proxies, but rather for the Fornax cluster. Some of the important results from our disc-break analysis is that the average 
($g-i$) colour of the outer disc i.e., beyond the break radius of LTGs, depends on their disc-break type. For galaxies detected in CO(1-0), if star formation differences created their disc-breaks, then molecular gas within their primary break radius could define the break radius. 

Further investigation on the stellar population content of the discs in LTGs from the Fornax3D data \citep{Sarzi2018,Iodice2019} and infall time via phase-space analysis \citep[e.g.][]{Rhee2017}
could address the origin of the different color gradients in Type~II and Type~III in the Fornax cluster.
In a forthcoming paper, we provide a detailed comparison of structure and evolution of LTGs in the cluster core and LTGs in the infalling SW subgroup centered on Fornax A (M.A.Raj, H.-S.Su et al. in prep), which is a different environment mainly populated by LTGs \citep{Iodice2017}.

\begin{acknowledgements}
The authors are immensely grateful to the anonymous referee for the useful comments and suggestions that significantly improved this article.  \\
 This publication has received funding from the European Union Horizon 2020 research and innovation programme under the Marie Sk\l odowska-Curie grant agreement n. 721463 to the SUNDIAL ITN network. \\

This work is based on visitor mode observations collected
at the European Organisation for Astronomical Research in the Southern
Hemisphere under the following VST GTO programs: 094.B-0512(B), 094.B-
0496(A), 096.B-0501(B), 096.B-0582(A).\\

 This research has made use of the NASA/IPAC Extragalactic Database (NED),
which is operated by the Jet Propulsion Laboratory, California Institute of Technology,
under contract with the National Aeronautics and Space Administration. \\

E.I and M. S. acknowledge financial support from the VST project (P.I. P. Schipani) \\
NRN acknowledges financial support from the “One hundred top talent program of Sun Yat-sen University” grant N. 71000-18841229.\\

GvdV acknowledges funding from the European Research Council (ERC) under the European Union's Horizon 2020 research and innovation programme under grant agreement No 724857 (Consolidator Grant ArcheoDyn).\\

We thank Dr. Ignacio Trujillo and Dr. Johan Knapen for their suggestions and comments on the break radius of LTGs, and  Dr. Crescenzo Tortora for his suggestions and comments. \\
We would also like to show our gratitude to Dr. Tim de Zeeuw for his insights on this work.

\end{acknowledgements}

 \bibliographystyle{aa.bst} % style aa.bst
  \bibliography{ltg} % your references Yourfile.bib
  %\bibliography{fornax}
% \onecolumn

\begin{appendix} 
%\twocolumn
\section{The Sample: Late Type Galaxies inside the virial radius} \label{ltgs}
In this section, we give a detailed description of the main properties of each galaxy analysed in this paper.

\subsection{FCC113}
FCC113, is an ScdIII pec, was classified as a star forming dwarf galaxy by \citet{Drinkwater2001}, but is a late type lopsided spiral galaxy with effective radius 1.98 kpc. Unlike the other regular late type galaxies studied, as part of this paper, FCC113 is located near the low-density region of galaxies in the virial radius of the Fornax cluster, at a projected distance of 1.209 deg. This galaxy, following on similar substructures, explained in the Sect. \ref{285_lit}. \citet{Drinkwater2001} point out to the ongoing star formation of this galaxy, which can be seen in the bright knots irregularly distributed in the northern regions of this galaxy. It appears to look like the galaxy is being pulled into the cluster centre, in the southern direction. 

\subsection{FCC115}
FCC115, an Sdm (edge on) galaxy, is the faintest galaxy in our sample. Due to the edge-on characteristics of this galaxy, it is difficult to discern the presence of spiral arms. It is located at a projected distance $D_{core}$ = 1.05 deg from the cluster centre, in the low-density regime. From image \ref{FCC115}, the SB image shows a dust lane in the centre, which is presumably the spiral arm (see Appendix. \ref{FCC115}). This galaxy has a  Type II profile, with a break radius of 1.47 kpc. The outer disc is bluer with the SB of 23.52 $mag/arcsec^2$ at the break radius.  

\subsection{FCC121}
Also known as the great barred spiral galaxy NGC 1365 has been studied in the past mostly concerning the supermassive black hole present in its core. The bar of this galaxy has dust and star formation that extends to the end of the spiral arms. With an effective radius of 12.74 kpc, this galaxy is located at a projected distance  $D_{core}$ = 1.06 deg from the cluster centre. It has a Type III profile, with a break radius of 18.46 kpc, and redder outer disc. 

\subsection{FCC176}
FCC176, also known as NGC 1369 is an SBa galaxy, consists of a bar and an outer ring formed by its spiral arms. It is located in the X-ray regions of  NGC 1399 in the central cluster. The effective radius of this galaxy is 2.68 kpc. It is at a projected distance $D_{core}$ = 0.82 deg from the cluster centre. The asymmetric halo (in the SE direction) and the outer ring which show intrinsic characteristics in the disc, suggest that this is an early type spiral galaxy \citep{Elmegreen1992, Mastropietro2005}, where the star formation in the outer regions appears to be stalled.
This galaxy has a Type II profile with a $B_r$ of 3.73 kpc. The outer disc is bluer with  average $g-i$ colour of 0.76 $\pm$ 0.2 mag (refer Tab. \ref{tab4}).The break radius is beyond the ring (OLR) of the galaxy at a SB of 22.20 $mag/arcsec^2$.  

\subsection{FCC179}
FCC179, also known as NGC 1386 is an Sa, Seyfert 2 galaxy, located at a projected distance $D_{core}$= 0.55 deg from the central galaxy, in the X ray regions of the Fornax cluster. It has an effective radius of 2.68 kpc.  This galaxy has been a topic of interest concerning its outflows \citep[e.g.][]{Rodriguez2017} and gas kinematics \citep[e.g.][]{Lena2015}. Being in the hot, high-density regime of the Fornax cluster, the spiral arms of this galaxy are concentrated in its central regions of 1 arcmin diameter. These spiral arms contains a lot of dust, which can be seen in its $g-i$ colour map at a level of 1.1 mag (see Appendix. \ref{FCC179}. This galaxy has a Type II profile, with a break radius of 7.01 kpc, that is $\sim$ 2.5 times the effective radius in $r$-band. 

\subsection{FCC263}
FCC263 is classified as a SBcd-III, barred spiral, located at $D_{core}$= 0.79 deg from the cluster centre. With an effective radius of 2.06 kpc, this galaxy is detected in HI  and CO(1-0), showing signs of ongoing star formation in its spiral arms \cite[see][]{Zabel2018,Schroder2001}. This galaxy has a Type III profile, with a break radius of 2.32 kpc.  
\citet{Zabel2018} point out to the irregular distribution of molecular gas, suggesting the possibility of tidal encounters or past minor mergers. Though we include this in our morphological segregation of regular spiral galaxies at smaller projected distance to the cluster centre, it is not clear if this galaxy is indeed within the X-ray halo or farther away, yet appears to have its spiral arms stripped. The outer isophotes are not aligned with the inner isophotes, and the galaxy is more thickened along the direction pointing towards the cluster centre. 

\subsection{FCC267}
FCC267, a Sm(IV) galaxy, is located at farthest projected distance (1.73 deg) from the cluster centre. This galaxy's spiral arms are concentrated in the innermost regions with double nucleus. It has an effective radius of 1.92 kpc. This galaxy has a Type III profile with a break radius of 1.43 kpc.

\subsection{FCC285} \label{285_lit}
FCC 285, also known as NGC 1437A , is an Sd-III galaxy. In the SB image, the galaxy's extended spiral arms are marked. This galaxy is detected in HI \citep{Schroder2001} and studied in FIR \citep{Fuller2014} and is located at $D_{core}$ = 1.17 deg from the cluster centre, with an effective radius of 4.90 kpc.  According to the De Vaucouleurs system of classification, Sd galaxies are usually diffuse, with a faint central bulge. These galaxies also have irregular star forming regions or star clusters spread out (see Appendix \ref{FCC285}). NGC 1437A has a similar arrow-shaped optical appearance as NGC 1427A and seems to be travelling in a southeast direction (based on the location of its own star-forming region) that is parallel with the orientation of NGC 1427A. The tail shows low SB regions, and the top part is elongated with the south, moving in opposite directions. NGC 1437A has about one third the HI mass as NGC 1427A and the velocity difference between these dIrrs is 1150 km s$^{-1}$, which is three times higher than the velocity dispersion of the Fornax cluster \citep{Lee-Waddell2018} with no indication of recent tidal interactions.

\subsection{FCC290}
FCC290, also known as NGC 1436 is an ScII galaxy at a projected distance $D_{core}$= 1.05 deg from the cluster centre, located within the X-ray regions of the central galaxy, with an effective radius of 4.64 kpc. The inner spiral arms have about a size of 2 arcmin consisting of dust lanes.
As the spiral arms are concentrated in the central regions (within 1R$_e$) of the galaxy, with a regular shape of disc in the outer regions, this galaxy appears to be moving to an S0 phase in its morphological evolution. This galaxy has a Type II profile with a break radius of 4.54 kpc. CO(1-0) detection  \citep{Zabel2018} shows presence of molecular gas in its spiral arms, within its break radius (see Appendix. \ref{FCC290}). 
\subsection{FCC302}
FCC302 is a diffuse, nearly edge-on galaxy with fairly uniform surface brightness, and possibly an irregular galaxy \citep{Matthews1997}, located at a projected distance $D_{core}$ = 1.305 deg. It has an effective radius of 2.76 kpc and a companion galaxy close to symmetric centre of this galaxy (NED). 

\subsection{FCC306}
FCC306, is a SBmIII, classified as a dwarf galaxy by \citet{Drinkwater2001}. It is the smallest galaxy of the sample we present, with an effective radius of 0.94 kpc. Though past research says that there is no evidence of substructure because of its small velocity dispersion in the Fornax cluster \citet{Drinkwater2001}, we speculate that there are spiral arms at a SB level of 29 mag in NW and SE directions marked in the contours of the SB image (see Appendix \ref{306_308}).   FCC306 lies in the beam of FCC308 and has been confirmed in a FLAIR observation with v = 915 $\pm15$ km s$^{-1}$ using emission lines \citep{Schroder1995}.

\subsection{FCC308}
FCC308, also known as NGC 1437b, is an Sd galaxy,  at a projected distance $D_{core}$ = 0.3 deg to the bright dwarf galaxy FCC306, with an effective radius of 4.46 kpc. The structure of this galaxy is similar to that of FCC312. Sd galaxies are usually edge-on, with ill-defined spiral arms \citep{Buta2011}.  One can say that the uneven distribution of star forming regions can be accounted for the spiral arms loosely wound, as marked in the $g-i$ colour map of this galaxy (see Appendix. \ref{FCC308}). There is a lot of dust present in the central regions of this galaxy, which can be seen in the $g-i$ colour map as well as the $g-i$ colour profile, where the dust extinction can be recognised out in the regions between 2-20 arcsec of the central galaxy. Flares on the outskirts of the disc is clear because of the deep images of FDS (see Appendix \ref{FCC308}).

\subsection{FCC312}
FCC312, is an Scd galaxy, with high stellar mass, detected in HI  \citep{Schroder2001} and CO(1-0) \citep{Zabel2018}, and studied in FIR (250 $\mu$m, \citealt{Fuller2014}). It has an effective radius of 11.30 kpc and has been a topic of interest in the past, concerning its structure, molecular clouds and regions of star formation. The central regions show the traces of ongoing star formation, which can be seen in the $g-i$ colour profile. It is located at a projected distance $D_{core}$ = 1.59  deg from the cluster centre. Being further away from the high-density regions of the central cluster, this galaxy shows the most extended disc  (boxiness) which is visible in the azimuthally averaged SB images shown in Appendix \ref{FCC312}. It has a Type III profile with a break radius of 6.91 kpc, around the same regions of star formation (see Appendix \ref{FCC312}). The boxy shape of edge-on spiral galaxies is said to be related to the vertical distribution of light \citep{Bureau1999}. Although past research  \citep[e.g][]{Bureau1999}  suggest that the presence of a boxy or peanut shaped bulge is due to the presence of a bar, this galaxy does not display a boxy bulge, but rather a boxy disc, with regions of high star formation surrounding the centre. The dust reddening of this galaxy makes the colour magnitude appear to be redder than bluer, and is clearly visible in the $g-i$ colour map of this galaxy (see Appendix \ref{FCC312}). The extended tail of the disc is at 25.6 $mag/arcsec^2$, which can be a sign of a recent minor merger event or due to instabilities caused in the disc.

%\onecolumn

\section{The Sample: Images and profiles}\label{sb_im}
In this section, we show the $g$-band VST images of LTGs inside the virial radius of the Fornax cluster, surface brightness profile in terms of effective radius, $g-i$ colour maps and ($g-i$) deconvolved (black) and original profiles (grey).
For 10 of the galaxies with disc-breaks in our sample, we also mark the break radius on the $g$-band images in surface brightness (black dashed lines) and  ($g-i$) colour profiles (red dashed lines). For 6 galaxies that were detected in CO(1-0), we plot the CO contours\citep{Zabel2018} in white.
 %NRN: right panel does not show the blue points crossing the red/green/purple?
 \begin{figure*}
  \centering
   \includegraphics[width=\hsize]{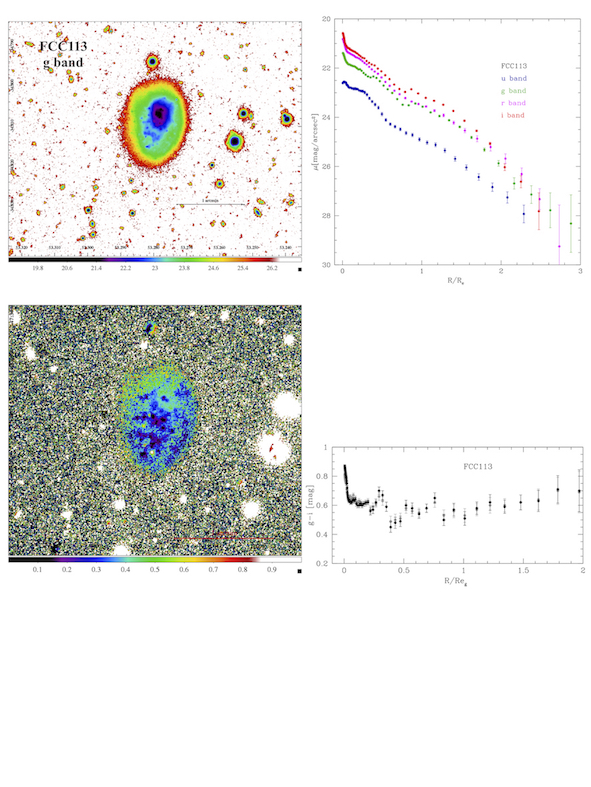}
     \vspace{-6.5cm}
     
      \caption{Surface Photometry of FCC113}
               
         \label{FCC113}
   \end{figure*}

\begin{figure*}
 \centering
  \includegraphics[width=\hsize]{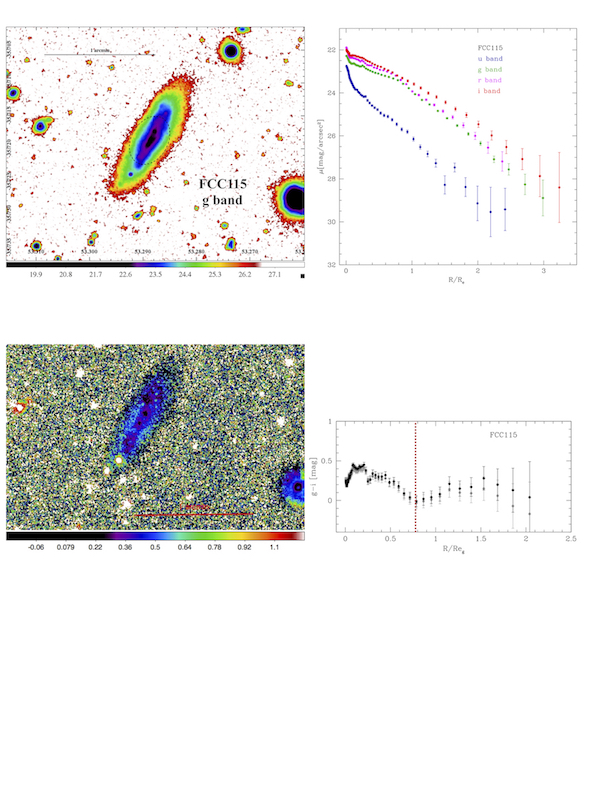}
    \vspace{-6.5cm}
     
      \caption{Surface Photometry of FCC115. SB image: Black dashed lines represent the break radius (Type II). ($g-i$) colour profile: Red dashed lines represent the break radius normalised to $Re_g$.}
             
         \label{FCC115}
   \end{figure*}

\begin{figure*}
  \centering
   \includegraphics[width=\hsize]{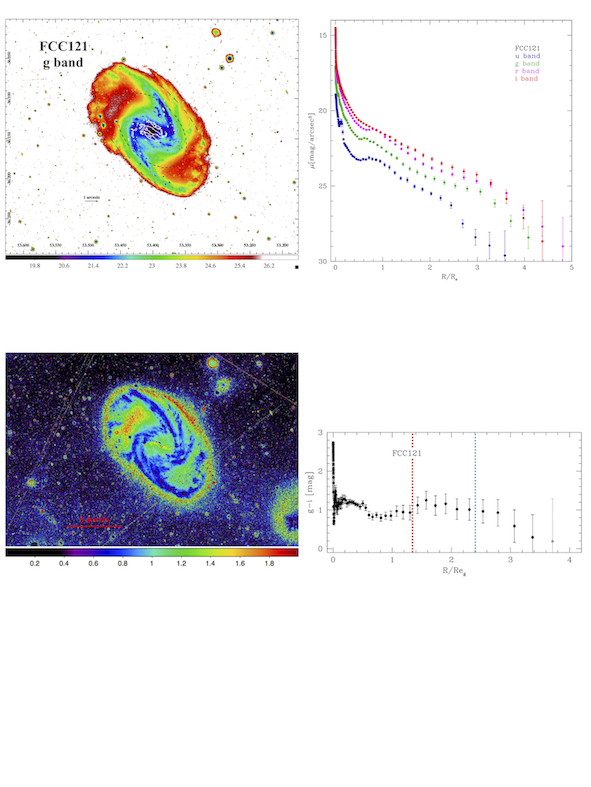}
     \vspace{-6.5cm}
     
      \caption{Surface Photometry of FCC121. SB image: Black dashed lines represent the inner break radius (Type III) and outer break radius (Type II). CO(1-0) contours are in white. ($g-i$) colour profile: Red dashed lines and cyan dashed lines represent the inner and outer break radii normalised to $Re_g$.}
             
         \label{FCC121}
   \end{figure*}

\begin{figure*}
  \centering
   \includegraphics[width=\hsize]{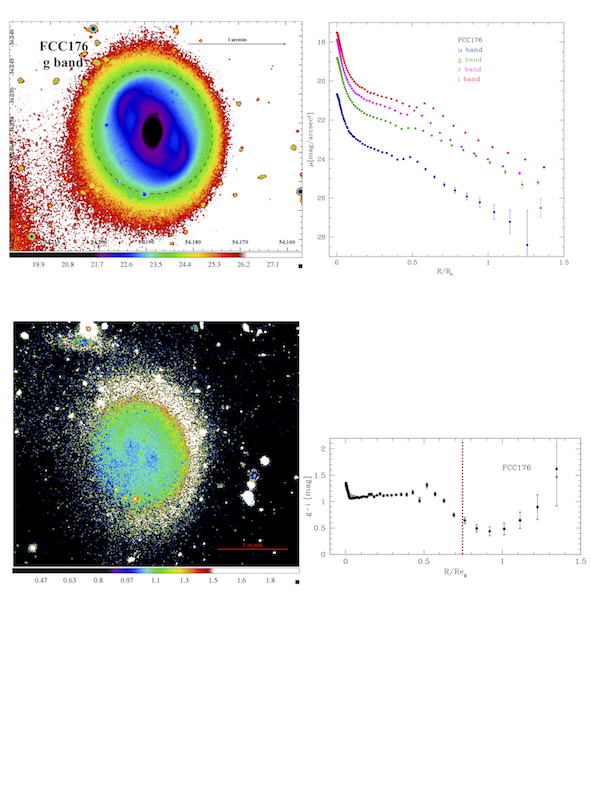}
       \vspace{-6.5cm}

      \caption{Surface Photometry of FCC176. SB image: Black dashed lines represent the break radius (Type II). ($g-i$) colour profile: Red dashed lines represent the break radius normalised to $Re_g$.}
             
         \label{FCC176}
   \end{figure*}
   
 \begin{figure*}
   \centering
   \includegraphics[width=\hsize]{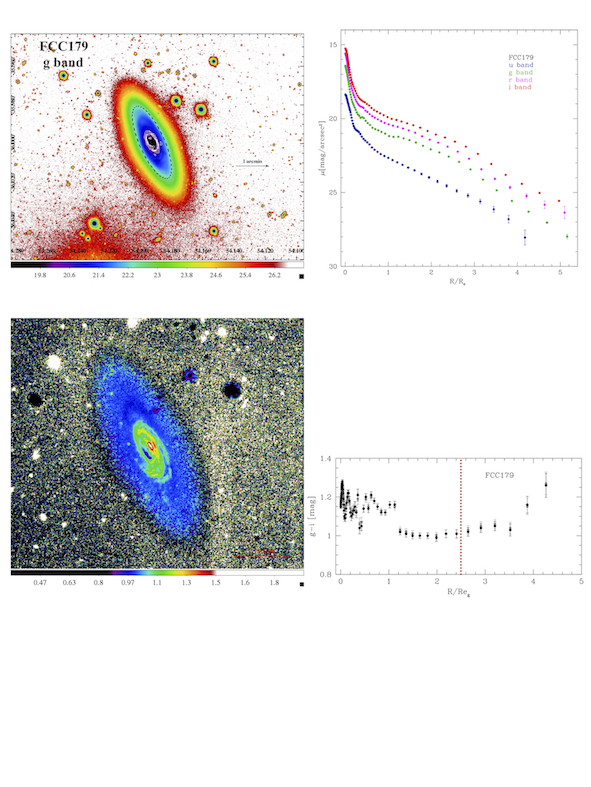}
    \vspace{-6.5cm}
    
      \caption{Surface Photometry of FCC179. SB image: Black dashed lines represent the  break radius (Type II). CO(1-0) contours are in white. ($g-i$) colour profile: Red dashed lines represent the break radius normalised to $Re_g$.}
       \label{FCC179}
      \end{figure*}

 \begin{figure*}
   \centering
   \includegraphics[width=\hsize]{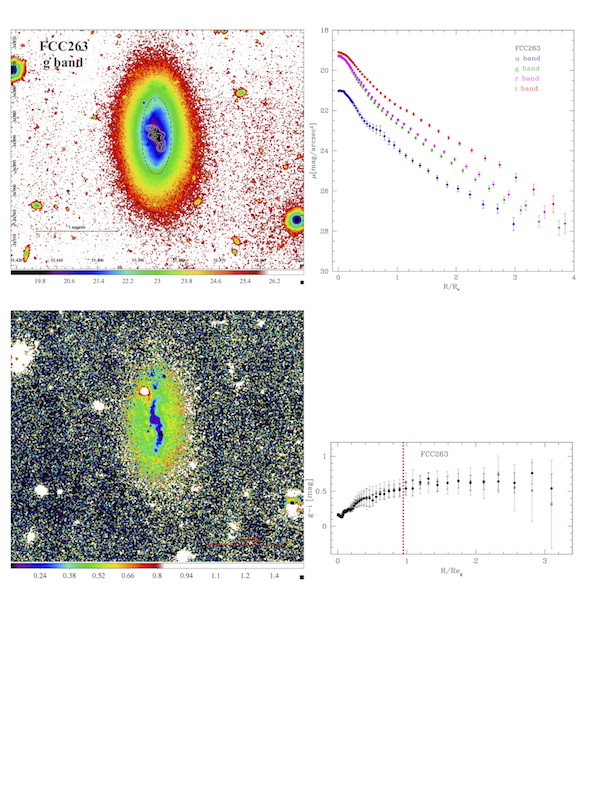}
   \vspace{-6.5cm}

      \caption{Surface Photometry of FCC263. SB image: Black dashed lines represent the break radius (Type III). CO(1-0) contours are in white. ($g-i$) colour profile: Red dashed lines represent the break radius normalised to $Re_g$.}
     \label{FCC263}
   \end{figure*}

\begin{figure*}
   \centering
   \includegraphics[width=\hsize]{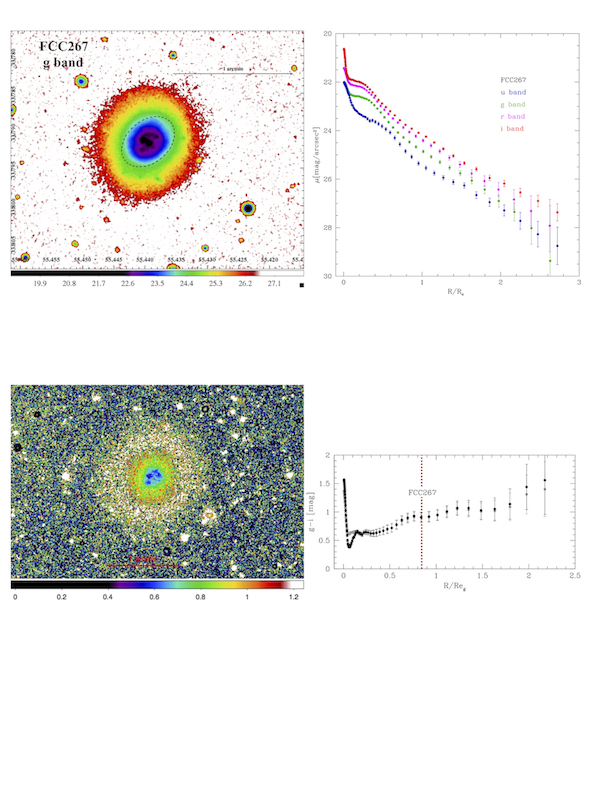}
    \vspace{-5.5cm}

      \caption{Surface Photometry of FCC267. SB image: Black dashed lines represent the break radius (Type III). ($g-i$) colour profile: Red dashed lines represent the break radius normalised to $Re_g$.}
             
         \label{FCC267}
   \end{figure*}

 \begin{figure*}
   \centering
   \includegraphics[width=\hsize]{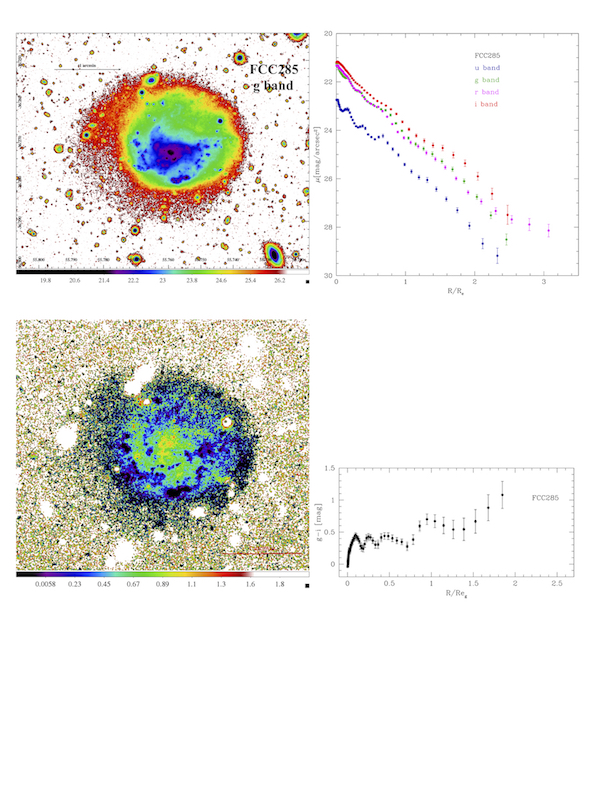}
    \vspace{-6.5cm}
    
      \caption{ Surface Photometry of FCC285.} 
      \label{FCC285}
   \end{figure*}

     \begin{figure*}
   \centering
   \includegraphics[width=\hsize]{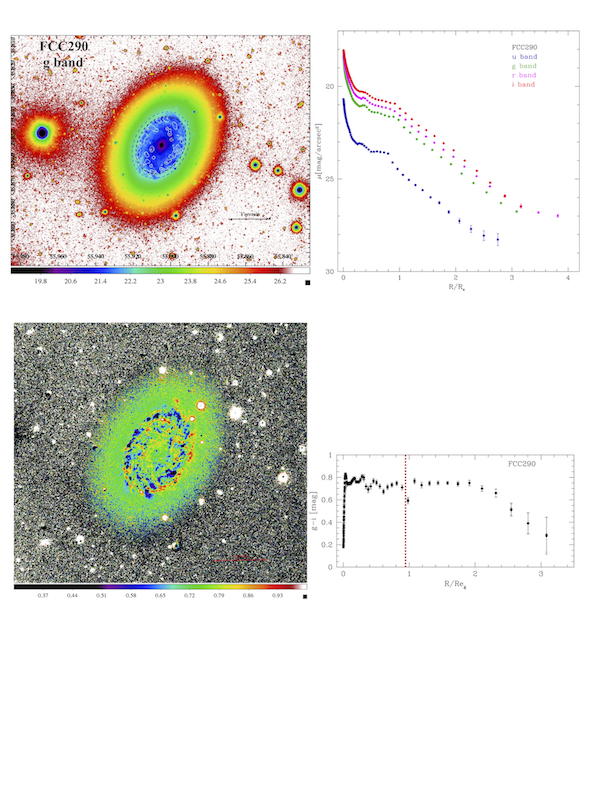}
    \vspace{-6 cm}
    
      \caption{Surface Photometry of FCC290. SB image: Black dashed lines represent the break radius (Type II). Blue dashed lines represent $B_r+ 2\sigma$.  CO(1-0) contours are in white. ($g-i$) colour profile: Red dashed lines represent the break radius normalised to $Re_g$.}

         \label{FCC290}
   \end{figure*}

  \begin{figure*}
   \centering
   \includegraphics[width=\hsize]{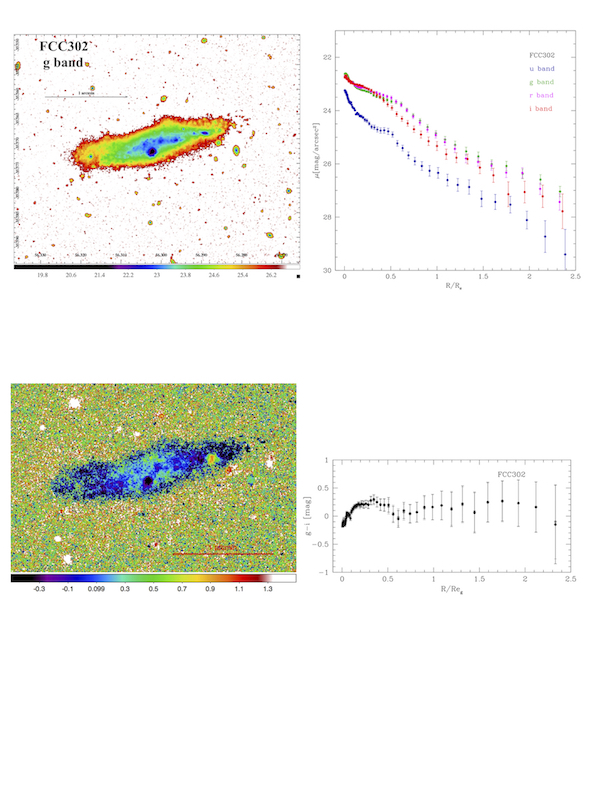}
    \vspace{-5.5cm}
    
      \caption{Surface Photometry of FCC302.}
             
         \label{FCC302}
   \end{figure*}

   \begin{figure*}
   \centering
   \includegraphics[width=\hsize]{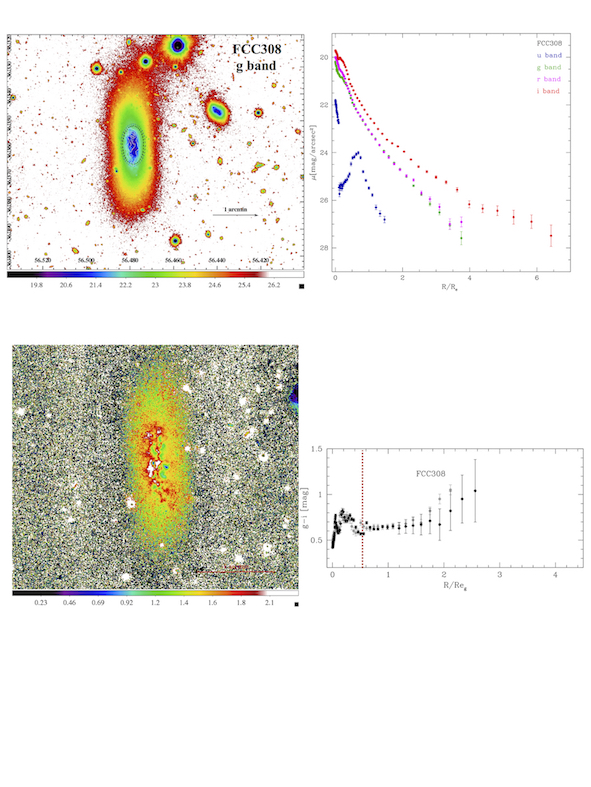}
    \vspace{-5 cm}

      \caption{Surface Photometry of FCC308. SB image: Black dashed lines represent the break radius (Type III) and blue dashed lines represent $B_r+ 2\sigma$. CO(1-0) contours are in white. ($g-i$) colour profile: Red dashed lines represent the break radius normalised to $Re_g$.}
             
         \label{FCC308}
   \end{figure*}

 \begin{figure*}
   \centering
   \includegraphics[width=16 cm]{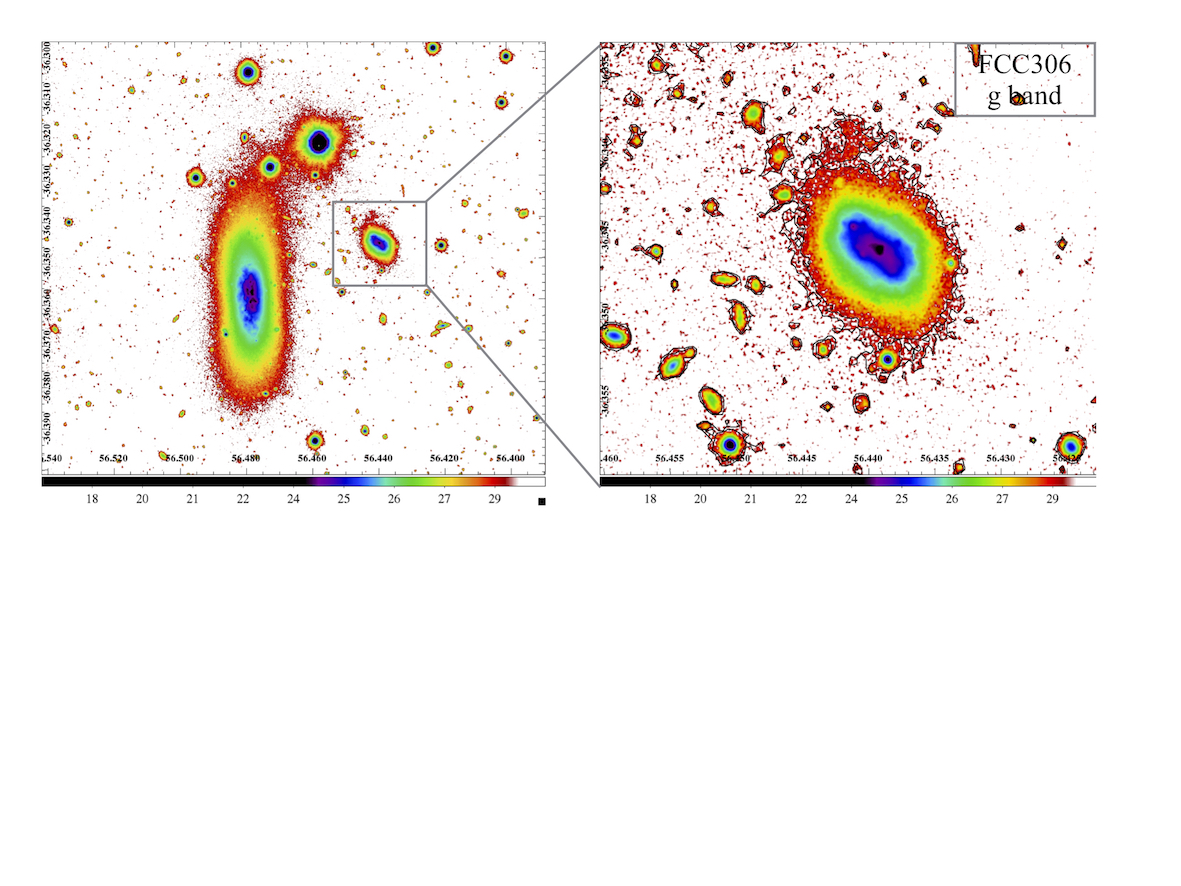}
    \vspace{-5 cm}
    
      \caption{FCC306 in zoom}
             
         \label{306_308}
   \end{figure*}

 \begin{figure*}
   \centering
   \includegraphics[width=16 cm]{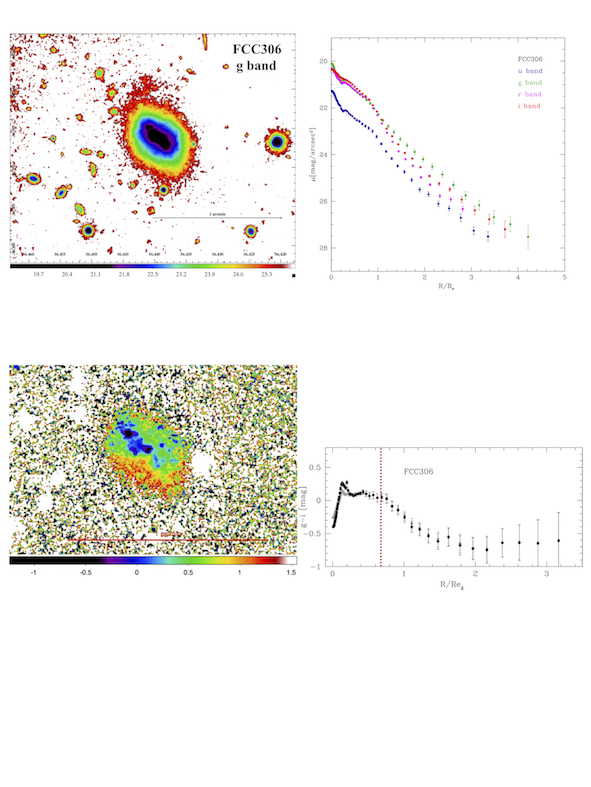}
    \vspace{-5 cm}
    
      \caption{Surface Photometry of FCC306. SB image: Black dashed lines represent the break radius (Type II). ($g-i$) colour profile: Red dashed lines represent the break radius normalised to $Re_g$.}
             
         \label{FCC306_mu}
   \end{figure*}

   \begin{figure*}
   \centering
   \includegraphics[width=\hsize]{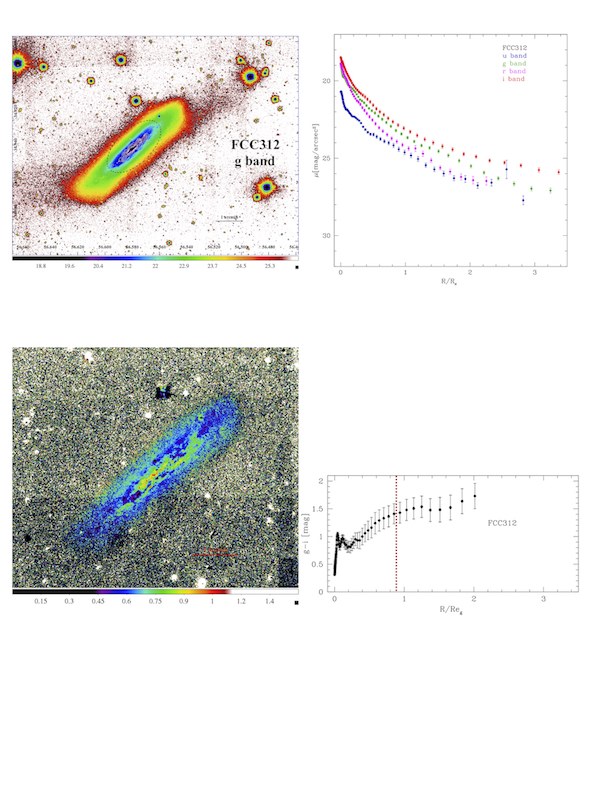}
    \vspace{-5 cm}
    
      \caption{Surface Photometry of FCC312. SB image: Black dashed lines represent the break radius (Type III). CO(1-0) contours are in white. ($g-i$) colour profile: Red dashed lines represent the break radius normalised to $Re_g$.}
             
         \label{FCC312}
   \end{figure*}

\section{Methodologies}\label{C}
\begin{figure*}
   \centering
   \includegraphics[width=17 cm]{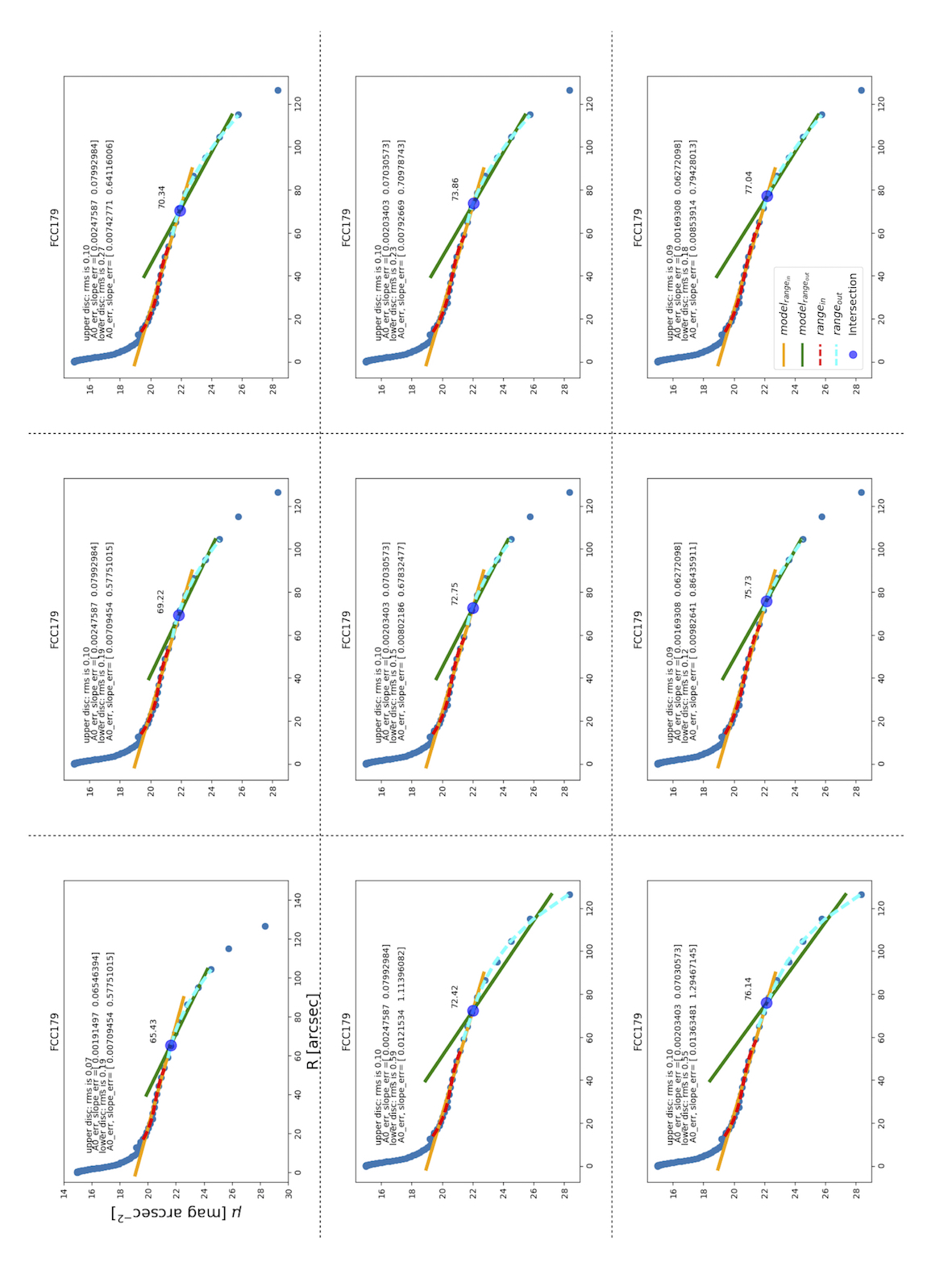}
    %\vspace{-5 cm}
    
      \caption{$(n+1)^3$ combinations of best fits on the inner and outer scale-lengths with minimal standard deviation, for FCC179 (here $n$ =2). $range_{in}$ is represented with a red spline, and $range_{out}$ is represented with a cyan spline. $model_{range_{in}}$ is marked in orange, while $model_{range_{out}}$ is marked in green. The break radius is marked at the intersecting point. Rms residuals, errors of the intercept, and slope of the fitted models for the inner and outer discs are mentioned in each plot. }
             
         \label{C1}
   \end{figure*}

\begin{figure*}
   \centering
   \includegraphics[width=17 cm]{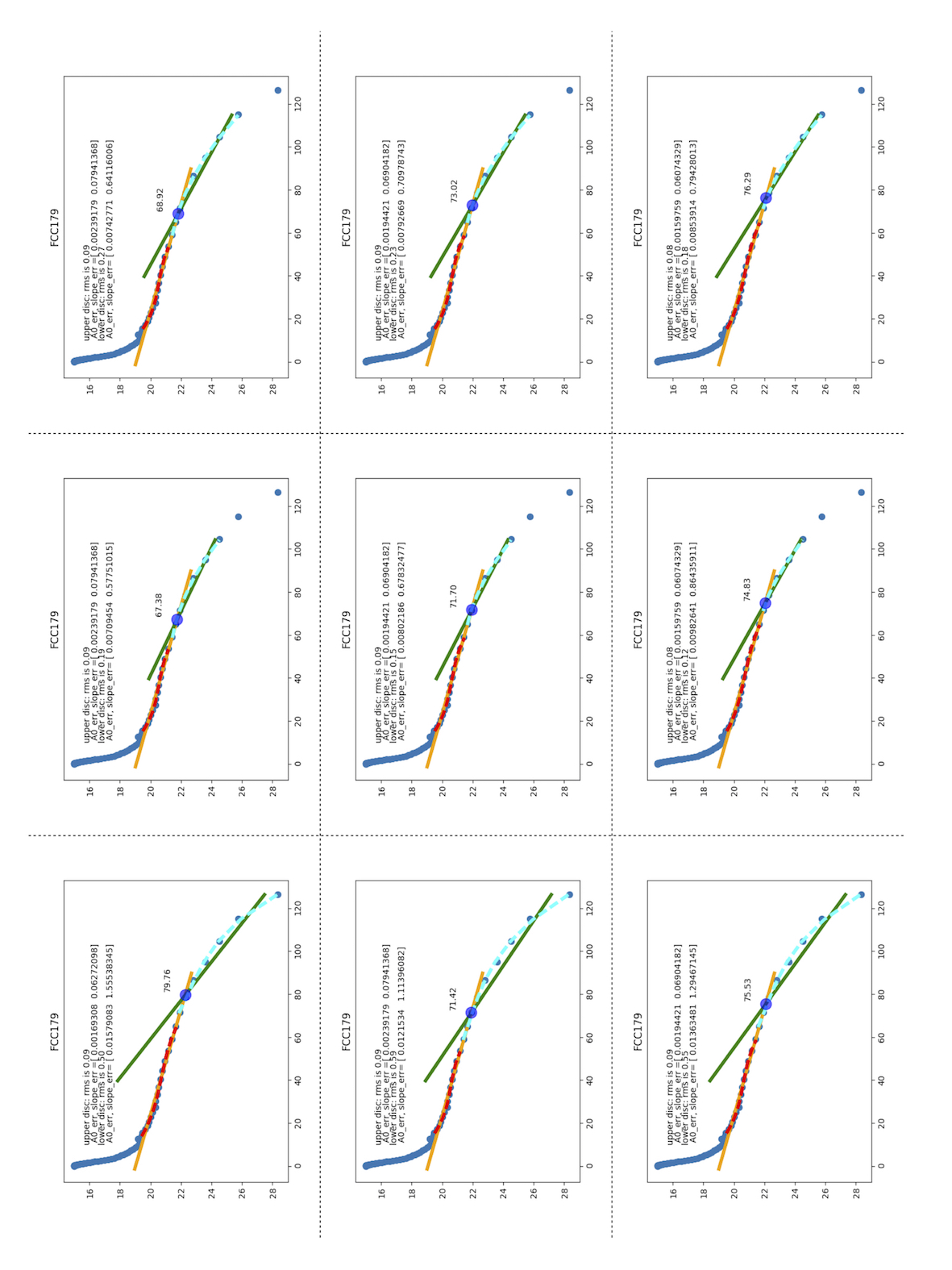}
    %\vspace{-5 cm}
    
      \caption{Appendix \ref{C1} continued.}
             
         \label{C2}
   \end{figure*}

\begin{figure*}
   \centering
   \includegraphics[width=17 cm]{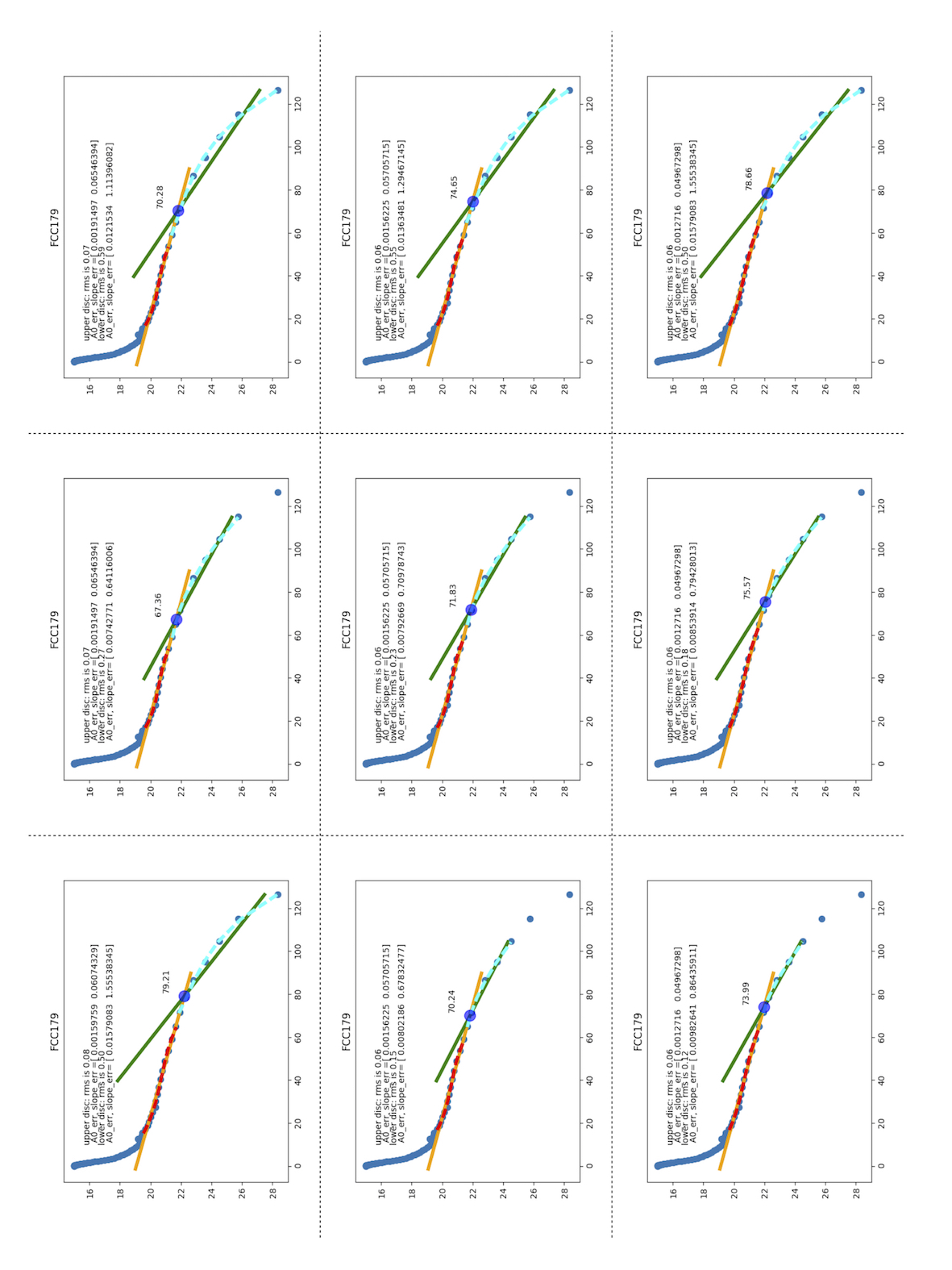}
    %\vspace{-5 cm}
    
      \caption{Appendix \ref{C1} continued.}
             
         \label{C3}
   \end{figure*}

Parameters of the best fit of  LTGs with disc breaks are given in Tab. \ref{tab5} with an example (FCC179) of the best fits produced by the algorithm shown in  Fig. \ref{C1}, \ref{C2}, \ref{C3}.

\begin{table*}[h!]
\caption{Parameters of the best fit of LTGs with disc breaks, inside the virial radius}              % title of Table
\label{tab5}      % is used to refer this table in the text
\centering                                      % used for centering table
\begin{tabular}{ccccc}          % centered columns (4 columns)

\hline\hline                        % inserts double horizontal lines
object 
& \multicolumn{2}{c}{$  h_{in} $ } 
& \multicolumn{2}{c}{$ h_{out}$ }
\\[+0.1cm]
 &\multicolumn{1}{c}{slope}
 &\multicolumn{1}{c}{${rms \ residuals}$} 
 &\multicolumn{1}{c}{slope}
 & \multicolumn{1}{c}{${rms \ residuals}$}
\\[+0.1cm]
  (1)
  & \multicolumn{2}{c}{(2)}
  &\multicolumn{2}{c}{(3)}
 \\
 \hline\hline

FCC115   &   0.09 $\pm$ 0.03 & 0.02 & 0.11 $\pm$ 0.07 &  0.02\\

FCC121  &   0.02 $\pm$ 0.11& 0.02 & 0.10 $\pm$ 0.18 &  0.07\\

FCC176  &   0.04 $\pm$ 0.06 & 0.09 & 0.06 $\pm$ 0.08 &  0.05\\

FCC179  &   0.04 $\pm$ 0.07 & 0.08 & 0.08 $\pm$ 0.70 &  0.23\\

FCC263 &   0.12 $\pm$ 0.14 & 0.08 & 0.07 $\pm$ 0.02 &  0.02\\

FCC267 &   0.17 $\pm$ 0.03 & 0.02 & 0.91 $\pm$ 0.03 &  0.02\\
FCC290  &   0.03 $\pm$ 0.09 & 0.07 & 0.05 $\pm$ 0.10 &  0.08\\
FCC306  &   0.20 $\pm$ 0.01 & 0.01 & 0.27 $\pm$ 0.07 &  0.08\\
FCC308 &   0.10 $\pm$ 0.05 & 0.05 & 0.05 $\pm$ 0.08 &  0.11\\
FCC312  &   0.05 $\pm$ 0.04 & 0.04 & 0.02 $\pm$ 0.14 &  0.06\\

\hline
\end{tabular}
\tablefoot{{\it Col.1 -} LTGs with disc break; {\it Col.2 -} slope of the fitted median model $h_{in}$ and its corresponding $rms \ residuals$;  {\it Col.3 -} slope of the fitted median model $h_{out}$ and its corresponding $rms \ residuals$} 
\vspace{10 cm}
 \end{table*}

   
\subsection{Reliability test} \label{test_dec}
In order to test the reliability of the method of deconvolution we use, we adopt the method illustrated by \citet[][hereafter B+17]{Borlaff17}.

According to B+17, images are deconvolved using the following operation: 
\begin{equation}
    Residuals = Image \ raw - PSF \ast Model_{GALFIT} \\
    \end{equation}
\begin{equation}
Deconvolved \ image = Model_{GALFIT} + Residuals
\end{equation}

Where $ PSF \ast Model_{GALFIT}$ is the 2D model (with best-fitted parameters) convolved with the adopted PSF, obtained from GALFIT3.0 \citep{Peng2010}, and $Model_{GALFIT}$ is the 2D model obtained using the best fit parameters of the galaxy from $ PSF \ast Model_{GALFIT}$, i.e. the model of the galaxy without PSF convolution. 

\begin{figure*}
%\vspace{-2 cm}
 \centering
  \includegraphics[width=\hsize]{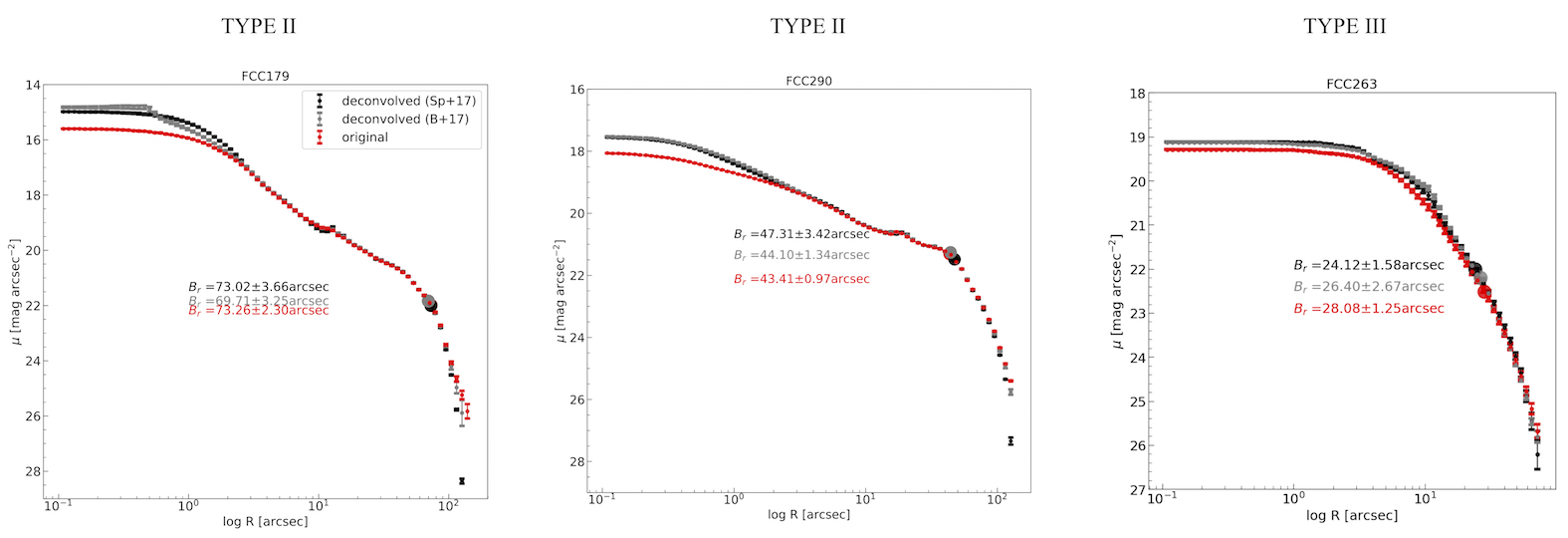}
      %\vspace{-3.5cm}

   \caption{Deconvolved SB profiles using SP+17 method (black) and the method by \citet{Borlaff17} (grey) over-plotted on the original (red) profiles for Type II (FCC179 and FCC290) and Type III (FCC263). }
   \label{galfit_test}
\end{figure*}

We apply this method to three of the galaxies in our sample: FCC179 and FCC290 (Type II with a second break) and FCC263 (Type III). For each of these galaxies, we use sersic+single exponential disc models to derive GALFIT3.0 models with PSF convolution. We fix the parameters obtained from  their $PSF \ast Model_{GALFIT}$ to extract another model, $Model_{GALFIT}$ i.e., without PSF convolution.  We then deconvolve the images using the above method. We extract the surface brightness profiles using \textsc{ellipse} and (see Sect. \ref{ana_dp}) compare it with the method (LR algorithm) given by SP+17 (see Fig. \ref{galfit_test}). 
We derive the break radius from the aforementioned radial profiles. From Fig. \ref{galfit_test}, it is clear that our method of deconvolution (LR algorithm) is consistent with the above method, and that deconvolution does not affect the location of the primary break radius (within the $\sigma_{nB_r}$) as well as the break-type, as the break radius we estimate is within the regions unaffected by the PSF. In the case of FCC179 and FCC290, a possible secondary break radius occurs in the regions that are affected by the PSF, where the contribution of the scattered light is accounted for with deconvolution.

   \begin{figure*}
   \centering
   \includegraphics[width=17 cm]{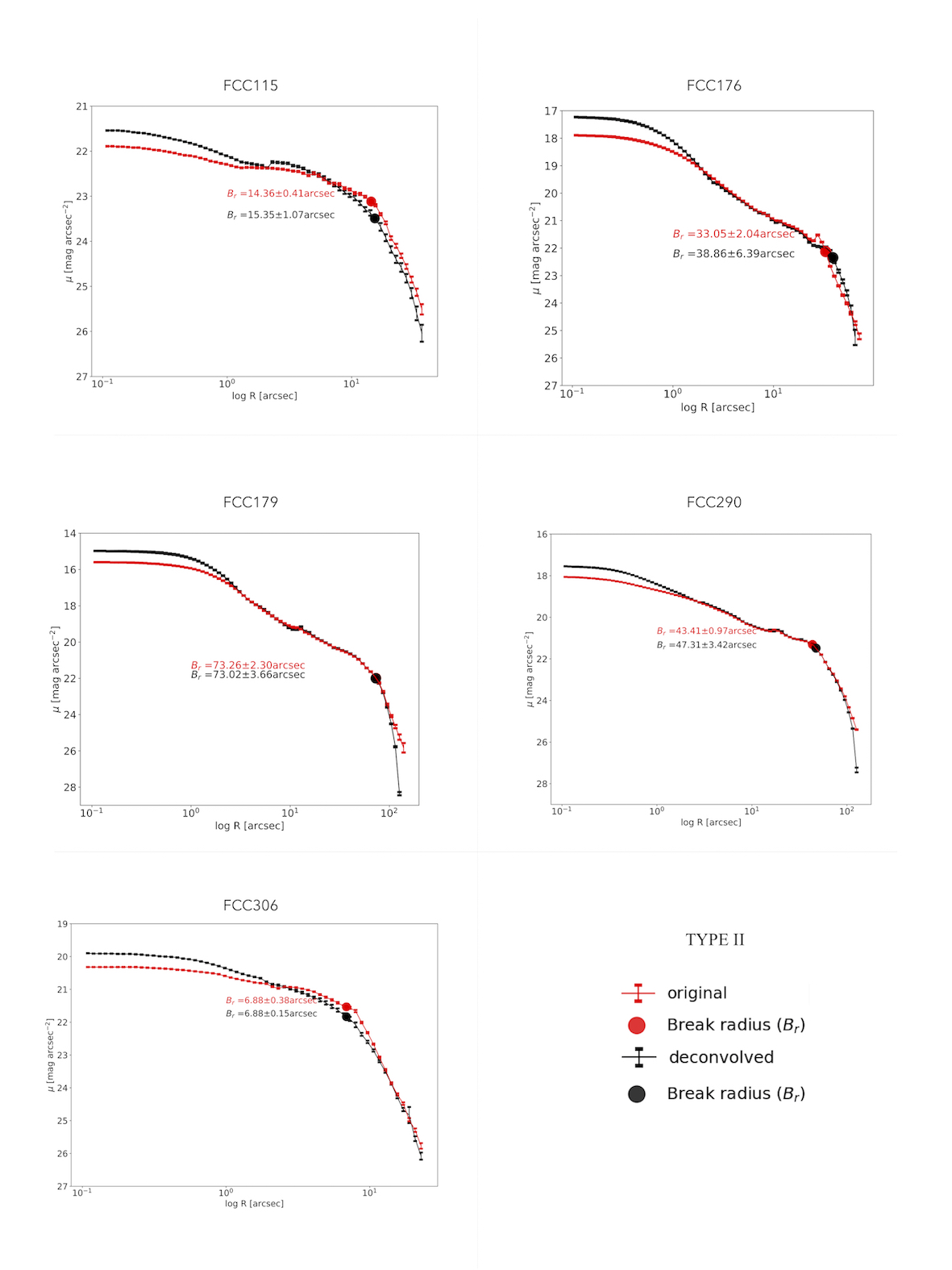} 
    %\vspace{-5 cm}
    
      \caption{Type II Deconvolved (black) and original (red) profiles in $r$-band with B$_r$ marked in their corresponding colours of each profile.}
             
         \label{dec_orig}
   \end{figure*}
  
   \begin{figure*}
   \centering
   \includegraphics[width=17 cm]{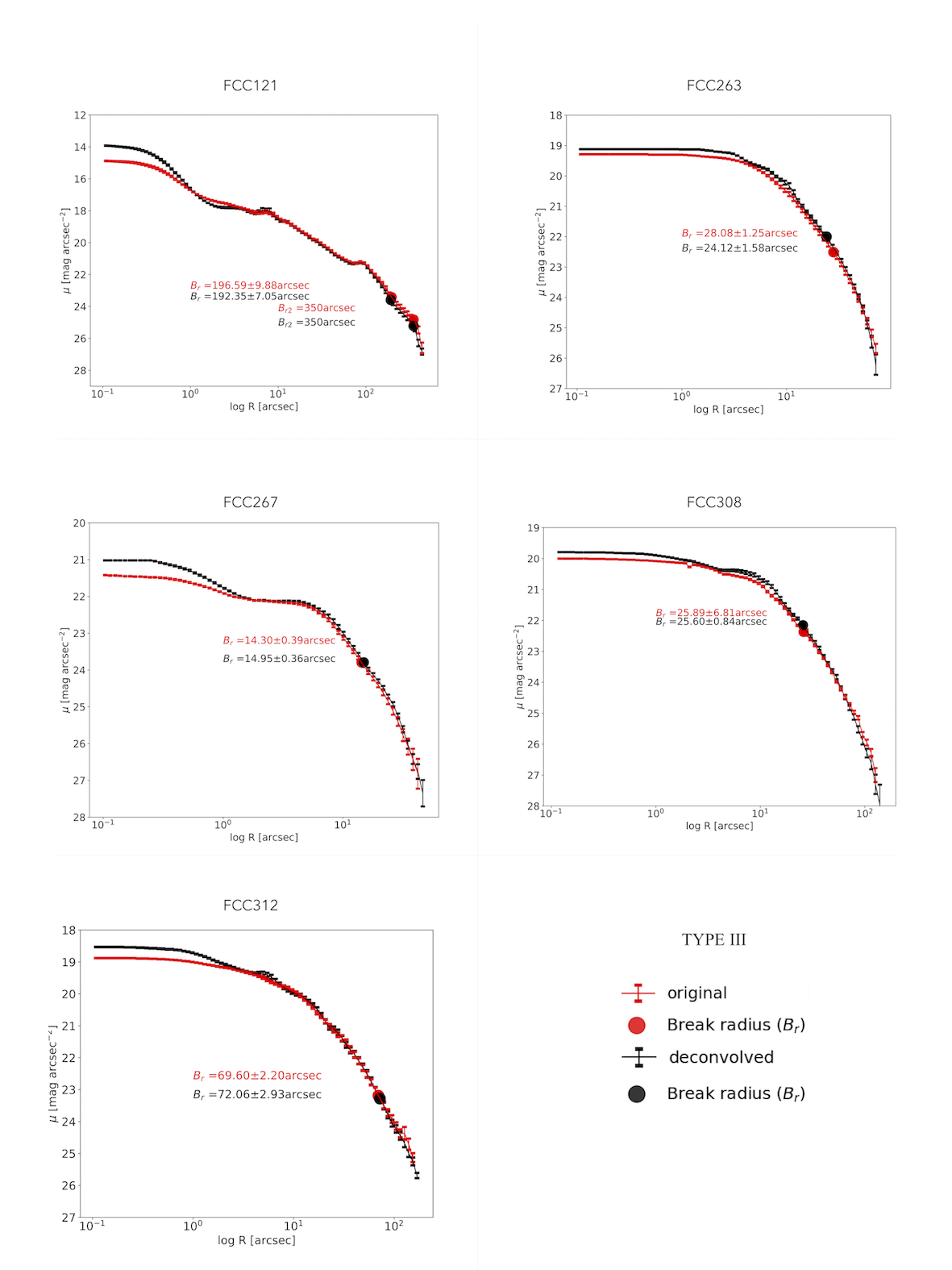}
    %\vspace{-5 cm}
    
      \caption{Type III (same as Fig. \ref{dec_orig})} 
             
         \label{dec_orig2}
   \end{figure*}
\end{appendix}
\end{document}